\documentclass[12pt,a4paper]{article}
 
\pdfoutput=1 
\usepackage{makecell}
\usepackage{graphicx}
\usepackage{epstopdf}
\usepackage{jcappub}
\usepackage{amsmath,amsthm,latexsym,amssymb,amsfonts,epsfig}
\usepackage{hyperref}
\usepackage{psfrag}
\usepackage[cp1250]{inputenc}
\usepackage[usenames,dvipsnames]{xcolor}
\usepackage{float}
\usepackage{multirow}
\usepackage{tabularx}
\newcolumntype{L}{>{\raggedright\arraybackslash}X}
\newcommand{\be}{\begin{equation}}
\newcommand{\ee}{\end{equation}}
\newcommand{\bea}{\begin{eqnarray}}
\newcommand{\eea}{\end{eqnarray}}
\numberwithin{equation}{section}

\title{Banks-Zaks Cosmology, Inflation, and the Big Bang Singularity}
\author{Michal Artymowski, Ido Ben-Dayan, Utkarsh Kumar}
\affiliation{ Physics Department, Ariel University, Ariel 40700, Israel}
\emailAdd{michal.artymowski@fuw.edu.pl}
\emailAdd{ido.bendayan@gmail.com}
\emailAdd{kumaru@ariel.ac.il}

\begin{abstract}
 {We consider the thermodynamical behavior of Banks-Zaks theory close to the conformal point in a cosmological setting. Due to the anomalous dimension, the resulting pressure and energy density deviate from that of radiation and result in various interesting cosmological scenarios. Specifically, for a given range of parameters one avoids the cosmological singularity. We provide a full "phase diagram" of possible Universe evolution for the given parameters. 
 For a certain range of parameters, the thermal averaged Banks-Zaks theory alone results in an  exponentially contracting universe followed by a non-singular bounce and an exponentially expanding universe, i.e. \textit{Inflation without a Big Bang singularity}, or shortly termed "dS Bounce". The temperature of such a universe is bounded from above and below. The result is a theory violating the classical Null Energy Condition (NEC). 
 Considering the Banks-Zaks theory with an additional perfect fluid, yields an even richer phase diagram that includes the standard Big Bang model, stable single "normal" bounce, dS Bounce  and stable cyclic solutions. 
 The bouncing and cyclic solutions are with no singularities, and the violation of the NEC happens only near the bounce. We also provide simple analytical conditions for the existence of these non-singular solutions. Hence, within effective field theory, we have a new alternative non-singular cosmology based on the anomalous dimension of Bank-Zaks theory that may include inflation and without resorting to scalar fields.}
\end{abstract}

\begin{document} 
\maketitle

\section{Introduction}
The existence of singularities in General Relativity (GR) signals the limitations of its validity as a theory describing Nature. Approaching the singularity, one expects some different classical or quantum theory to be the proper description of physical phenomena smoothing out the singularity. 
The striking difference between singularities in GR compared to singularities in other theories like electromagnetism {is stated by Wald in \cite{Wald:1984rg}, where he writes: {\it (\ldots) the ``big bang'' singularity of the Robertson-Walker solution is not considered to be part of the space-time manifold; it is not a ``place'' or a  ``time''. Similarly, only the region $r>0$ is incorporated into the Schwarzschild space-time; unlike the Coulomb solution in special relativity, the singularity at $r=0$ is not a ``place''.}}The most celebrated examples of singularities in GR are the black hole singularity and the Big Bang singularity in Cosmology. The singularity theorems \cite{Penrose:1964wq,Hawking:1966sx,Hawking:1966jv,Hawking:1967ju,Hawking:1969sw} are based on several assumptions. Most notably in the context of Cosmology is the existence of energy conditions. For a closed Universe with $k=+1$ the singularity occurs if the energy momentum tensor fulfills the so-called strong energy condition. For the open or flat Universe, $k=-1,0$ it is sufficient to assume the Null Energy Condition (NEC) \cite{Visser:1999de}:
\be
\theta_{ab}k^ak^b \geq 0
\ee
where $\theta_{ab}$ is the energy momentum tensor and $k^a$ is a future pointing null vector field.
If we can use a perfect fluid description, then the left hand side of the above inequality can be written in terms of the fluid's energy density $\rho$, its pressure $p$ and the 4-velocity $u_a$ resulting in: 
\be
\theta_{ab}=(\rho+p)u_au_b+p\,  g_{ab} \Rightarrow \rho+p \geq 0.
\ee
If the equation of state of the fluid is given by $p=w\rho$, then the NEC further simplifies to $w\geq -1$.
These energy conditions are not based on first principles, but rather on familiar forms of matter, and the fact that violating these conditions many times lead to instabilities invalidating the analysis.
Hence, it may very well be that the Universe had a phase (or phases) of NEC violation without ever reaching a singularity. These non-singular solutions are especially interesting if they are predictive with predictions that can be tested in cosmological settings such as CMB observations or Laser Interferometers. 
We are therefore interested in theories that violate the NEC, but still enjoy a stable evolution. A well known example of such theories are Galilean theories \cite{Qiu:2011cy} where non-canonical kinetic terms exist, providing a bouncing solution: The Universe turns from a contracting phase to an expanding one, the scale factor $a(t)$ is always finite, the Hubble parameter $H$  changes sign from negative to positive and exactly vanishes at the bounce \cite{Ijjas:2018qbo,Brandenberger:2016vhg,Battefeld:2014uga,Artymowski:2008sc,Artymowski:2010by,Artymowski:2013qua,Ben-Dayan:2018ksd,Ben-Dayan:2016iks}. Discussions on the validity and stability of these scenarios are discussed in \cite{Qiu:2011cy,Rubakov:2014jja,Dubovsky:2005xd,Qiu:2013eoa,Qiu:2015nha,Cai:2016thi,Cai:2017tku,Cai:2017dyi,Easson:2011zy,Libanov:2016kfc,Ijjas:2016vtq,Ijjas:2016tpn,Koehn:2015vvy,Battarra:2014tga,Sawicki:2012pz,Dobre:2017pnt}.\footnote {{In brief, in the context of scalar field theories it was shown, for example in \cite{Rubakov:2014jja,Dubovsky:2005xd}, that the NEC violation may in particular lead to tachyonic, gradient, and ghost instabilities. As a consequence there is a rapid uncontrolled growth of perturbations. If this situation persists for a long enough time, the analysis looses its validity. Another peculiarity is that NEC violation, implies there is an observer seeing arbitrarily negative energy densities and a hamiltonian unbounded from below \cite{Sawicki:2012pz,Dobre:2017pnt}.} }

In this manuscript we take a different approach. NEC violation is well known to occur in QFT \cite{Witten:2018lha}, and do not seem to lead to any problems in the validity of the analysis. {Moreover, the energy density in QFT is {\emph not} positive semi-definite. So a violation of the NEC should not be considered as a cardinal no-go theorem. The space of QFTs is not limited to scalar fields, even with non-canonical kinetic terms. In this work we} deviate from the scalar field paradigm and would like to consider other field theories. For our purposes, we will consider the Banks-Zaks theory \cite{Banks:1981nn,Georgi:2007ek}. Banks-Zaks theory has a non-trivial IR conformal fixed point. At this point the energy-momentum tensor is traceless. If we slightly shift away from the fixed point the beta function does not vanish anymore and the trace of the energy momentum tensor will be proportional to the beta function. As a result, we can consider a thermal average of the theory yielding well-defined expressions for the energy density and pressure of the fluid as functions of temperature. This was done in \cite{Grzadkowski:2008xi}, and was dubbed "unparticles" even though the notion of unparticles may be more general. For ease of presentation and for the rest of the discussion, we will refer to the thermal average of the Banks-Zaks theory slightly removed from the fixed point as unparticles.
The resulting energy density and pressure deviate from that of radiation due to the anomalous {dimension} of the theory. The theory can violate the NEC yielding a very rich "phase diagram" of the possible cosmological behavior. 

Starting from the unparticles only scenario, we find a new regular bouncing solution on top of the existing ones. The Universe exponentially contracts, bounces and exponentially expands, i.e. reaching an inflationary phase, henceforth "dS Bounce". The temperature is bounded and is minimal at the bounce and maximal at the exponential contraction/expansion. The NEC is always violated. We then consider the existence of an additional perfect fluid. Depending on the parameters of the theory we have the standard singular solutions, cyclic solutions, and single bouncing solutions including the dS Bounce or a "normal bounce" that is preceded by a slow contraction and followed by a decelerated expansion. 
The NEC is only violated around the bounce.
Finally, the analysis in \cite{Grzadkowski:2008xi} identified the normalization scale and the temperature $\mu=T$. We wish to consider a different setting where the normalization point $\mu$ is fixed, unlike the temperature that can vary, hence, $\mu\neq T$. This will greatly constrain the allowed anomalous dimension of the theory, which will generate a somewhat different phase diagram. The outcome is a new class of viable non-singular cosmological models based on the anomalous dimension of a gauge theory rather than a scalar field. 

The paper is organized as follows. We start with reviewing the analysis of \cite{Grzadkowski:2008xi} in section \ref{thermo}. We then consider a Universe filled only with unparticles and its phase diagram in section \ref{sec:uonly}. In section \ref{sec:fluid} we analyze the case of unparticles with additional perfect fluid. We reassess the constrained case of $\mu\neq T$ in section \ref{sec:fixedmu}. We then conclude, identifying novel research directions. A brief discussion of the singular solutions in our scenario is relegated to an appendix.


\section{Thermodynamics of Banks-Zaks theory}\label{thermo}
In this section we review the results of \cite{Grzadkowski:2008xi}. In order to study the thermodynamic behavior of unparticles, one assumes that the trace of the energy momentum tensor ($ \theta^{\mu}_{\mu}  $) of a gauge theory where all the renormalized masses vanish \cite{Collins:1976yq} as
\begin{equation}
\theta_\mu^\mu = \frac{\beta}{2 g} N \left[ F_{a}^{\mu \nu}  F_{a \,\mu \nu}\right]\, ,
\label{eq:trace}
\end{equation}
where $ \beta $ denotes the beta function for the coupling $g$ and $N$ stands for a normal product. For unparticles the $\beta $ function has a non-trivial IR fixed point at $ g = g_* \neq 0 $. Close to the fixed point one finds 
\begin{equation}
 \beta = a\left( g -g_*\right), \qquad a>0 \, .
\end{equation}
In such a case the running coupling reads 
\begin{equation}
g(\mu) = g_* + u \mu^a; \qquad \beta[g(\mu)] = au\mu^a \, , \label{g}
\end{equation}
where $u$ and $ \mu$ are integration constant and renormalization scale respectively. We are interested in lowest-order corrections to the conformal limit (where $ \theta_{\mu}^{\mu} = 0$ ) where the system is in thermal equilibrium at the temperature T and does not contain any net conserved charge. As pointed out earlier $\beta $ vanishes in the conformal limit, so $ \langle N\left[ F_a^{\mu} F_{a \, \mu \nu}\right]\rangle$ is equal to its conformal value. Performing a thermal average and taking the renormalization scale $ \mu = T $, one finds 
\begin{equation}
\langle N\left[ F_a^{\mu} F_{a \, \mu \nu}\right]\rangle = b T^{4 + \gamma } , \label{ev}
\end{equation}
where $\gamma $ is the anomalous dimension of operator. Using the trace equation ($  \langle \theta^{\mu}_{\mu} \rangle = \rho_u -3p_u$) along with equations (\ref{g}) and (\ref{ev}) results in
\begin{equation}\label{eq:rhoptrace}
\rho_u -3p_u = A T^{4 + \delta}; \quad\left( A = \frac{a u b}{2 g_*} \,,\, \delta = a +\gamma \right) ,
\end{equation}
where $ \rho_u $ and $ p_u$ are the energy density and pressure of unparticles.  

By using the first law of thermodynamics $d(\rho V ) + p dV = T d(sV )$ gives the energy density, the pressure and the entropy density of the Banks-Zaks fluid as functions of temperature \footnote{Since any $\delta$ can be compensated by changing $A$ that has an arbitrary integration constant, we define $B\equiv A \left( 1+\frac{3}{\delta}\right)$ for the purpose of simplifying our analysis. The special cases where we have to return to \eqref{eq:rhoptrace}, $\delta \rightarrow -3,0$ are presented below. These limiting cases are regular and well behaved.}:
\begin{eqnarray}
\rho_u &=& \sigma T^4 + A \left( 1+\frac{3}{\delta} \right) T^{4+\delta} \equiv \sigma T^4 + B \,T^{4+\delta} \, , \label{eq:rhou} \\
p_u &=&  \frac{1}{3}\sigma T^4 +\frac{A}{\delta} T^{4+\delta} \equiv \frac{1}{3}\sigma T^4 +\frac{B}{3+\delta} T^{4+\delta} \, . \label{eq:pu} \, ,\\
s_{u}&=& \frac{4}{3} \sigma T^3 + A\left( 1 + \frac{4}{\delta}\right)T^{3+\delta} \equiv \frac{4}{3} \sigma T^3 + B\left( \frac{4+\delta}{3+\delta}\right)T^{3+\delta}  \, ,
\end{eqnarray}
where $ \sigma $ is a positive integration constant, related to the number of relativistic degrees of freedom, and $s_u$ stands for the entropy density of unparticles.  
$B=0$ corresponds to the standard radiation case, while the limit $\delta\rightarrow0$ corresponds to logarithmic corrections to radiation in the form 
of $\rho_u=\sigma T^4+3AT^4\ln T,\, p_u=(\sigma-A)/3T^4+AT^4\log T$ and $\delta\rightarrow-3$ corresponds to $\rho_u=\sigma T^4,\, p_u=1/3\left(\sigma T^4-A T\right)$.

{The analysis presented above is valid under two assumptions. First of all, unparticles are just an effective theory of Banks-Zaks valid below the scale $\Lambda_\mathcal{U}$. Thus, we assume that $T < \Lambda_\mathcal{U}$ throughout the whole evolution of the Universe. Note that in the $T\gg \Lambda_\mathcal{U}$ one restores asymptotically the free Banks-Zaks theory with $\rho =\sigma_{BZ} T^4$, where $\sigma_{BZ}\gg 1$.} {In such a regime Banks-Zaks and standard model particles are coupled and this coupling is in fact a source of the anomalous dimension of the Banks-Zaks sector. However, for  $T<\Lambda_\mathcal{U}$ Banks-Zaks sector effectively decouples from the standard model and therefore throughout the paper we will assume that unparticles and any other additional matter are decoupled.} 

{Second, in order to use the first law of thermodynamics in the above form, one must assume adiabatic changes of temperature. In a system without dissipative effects (which is true in our case, since we assume $T<\Lambda_\mathcal{U}$) one expects the evolution of the Universe to be adiabatic. In \cite{Fukuma:2003kv} the authors have presented the alternative approach, in which adiabaticity requires satisfying an additional condition $\omega > H$, where $\omega\sim T$ is the frequency of unparticles. As we will show, this requirement is fulfilled in our case. Hence, in both cases the evolution of the Universe is adiabatic even near the bounce.}


\section{Bouncing solution with unparticles}
\label{sec:uonly}
Let us assume that the Universe is filled with unparticles with temperature $T$ and with the energy density $\rho_u$ and pressure $p_u$ of equations \eqref{eq:rhou} and \eqref{eq:pu}.  
Assuming the flat FLRW metric, the Friedmann equations are: 
\begin{eqnarray}
3H^2 &=& \rho \, , \label{eq:Fried1} \\
\dot{H} &=& -\frac{1}{2} (\rho+p) \, , \label{eq:Fried2}
\end{eqnarray}
with $\rho=\rho_u$ and $\ = p_u$. We define two "extremal" temperatures that will turn out to be crucial in understanding the evolution of the Universe.
First, the bounce temperature, which is the solution to $H=\rho=0$:
\begin{equation}
T_b = \left(-\frac{\sigma}{B} \right)^{\frac{1}{\delta}}  \, .\label{eq:T0pure}
\end{equation}
Second, the "critical temperature", at which $\dot{H}=0$:
\be
T_c=\left[\frac{4(\delta+3)}{3(\delta+4)}\left(-\frac{\sigma}{B} \right)\right]^{\frac{1}{\delta}}=\left[\frac{4(\delta+3)}{3(\delta+4)}\right]^{\frac{1}{\delta}}T_b
\ee
Throughout the manuscript we will many times prefer to work with dimensionless quantities, so we define $x\equiv T/T_b$ and $y\equiv T/T_c$.

Our goal is to obtain a bounce, which is possible only if at a certain time $t_b$ one can satisfy the following conditions
\begin{equation}
H_b = H(t_b) = 0 \, , \qquad \dot{H}_b = \dot{H}(t_b) > 0 \, , \label{eq:bounceconditions}
\end{equation}
where the subscript $b$ corresponds to the value at the bounce. From Eqs. (\ref{eq:Fried1},\ref{eq:Fried2}) one finds the following conditions for the existence of the bounce
\begin{equation}
\rho_{b} = 0\, , \qquad p_{b} < 0 \, . \label{eq:bouncecondition}
\end{equation}
By definition $\sigma > 0$, so in order to obtain $\rho_u = 0$ one requires
\begin{equation}
B < 0  \, .
\end{equation}
The pressure at the bounce is equal to $p_{ub} = \frac{\delta  \sigma  T_b^4}{3(\delta +3)}$ which means that in order to obtain $p_{ub} < 0$ one requires $\delta \in (-3,0]$. We rewrite the energy density and pressure in terms of $x$:
\begin{equation}
\rho_u = \sigma T_b^4 x^4(1-x^\delta) \, , \qquad p_u =  \sigma T_b^4 x^4 \left(\frac{1}{3}-\frac{x^{\delta }}{\delta +3}\right) \, .
\end{equation}
For consistency the continuity equation must hold and can be written as
\begin{equation}
\dot{\rho}_u + 3H (\rho_u+p_u) = \frac{dx}{dt} \left(\frac{d \rho_u}{dx} + 3\frac{a_x}{a}(\rho_u+p_u)\right) = 0 \, \label{eq:contu} 
\end{equation}
where $a_x=\frac{da}{dx}$.  

\subsection{Solutions for $ \delta \in (-3\,,\,0]$}

Equation \eqref{eq:contu} can be solved analytically, which for $\delta \neq -3$ gives\footnote{Let us assume for the moment that $T$ has no lower bound. Then the scale factor has 3 possible limits for $T \to 0$. For $\delta > -3$ one finds $a\to \infty$, which corresponds to the future infinity in the expanding and cooling Universe. For $\delta<-3$ one finds $a\to 0$, which corresponds to $\rho_u\to\infty$ and the Big Bang singularity (for $\delta<-4$) or to $\rho_u\to 0$ and Minkowski-like initial conditions (for $-4<\delta<-3$). The $\delta =- 3$ is unique, since it is the only case, for which $a \to 1$ and $\rho_u \to 0$.}:

\begin{equation}
a = \frac{1}{x} \left(\frac{-\delta }{3  (\delta +4) x^{\delta }  -4(\delta +3)}\right)^{1/3} \ {\propto \ y^{-1}(y^{\delta}-1)^{-1/3}} \, . \, . 
\label{eq:scale}
\end{equation}
We normalize the scale factor to be unity at the bounce, $T_b$, $(x=1)$. The scale factor has a pole at the critical temperature  $T= T_c$. Due to the existence of the pole the temperature must remain bigger or smaller than $T_c$ throughout all of the evolution. We are interested in a bouncing Universe and therefore we choose $T<T_c$. We want to emphasize that the existence of the pole of $a$ does not mean that the theory has any singularity. Both, curvature and energy density, are finite at $T=T_c$. The $a\to \infty$ limit simply corresponds to $t\to \infty$.\footnote{Note that the $T>T_c$ would require a different normalization of $a$ to keep $a$ real and positive. In such a range of $T$ solutions asymptote to dS space with $T=T_c$ from above.} 
Note that for $B = 0$ one recovers $a \propto 1/T$. From \eqref{eq:scale} one finds:

\begin{equation}
H = -\frac{(\delta +3) \left((\delta +4) x^{\delta
   }-4\right)}{x \left(3 (\delta +4) x^{\delta
   }-4 (\delta +3)\right)}\frac{d x}{d t}\, .
\end{equation}

\begin{figure}
\centering
\includegraphics[height=4.3cm]{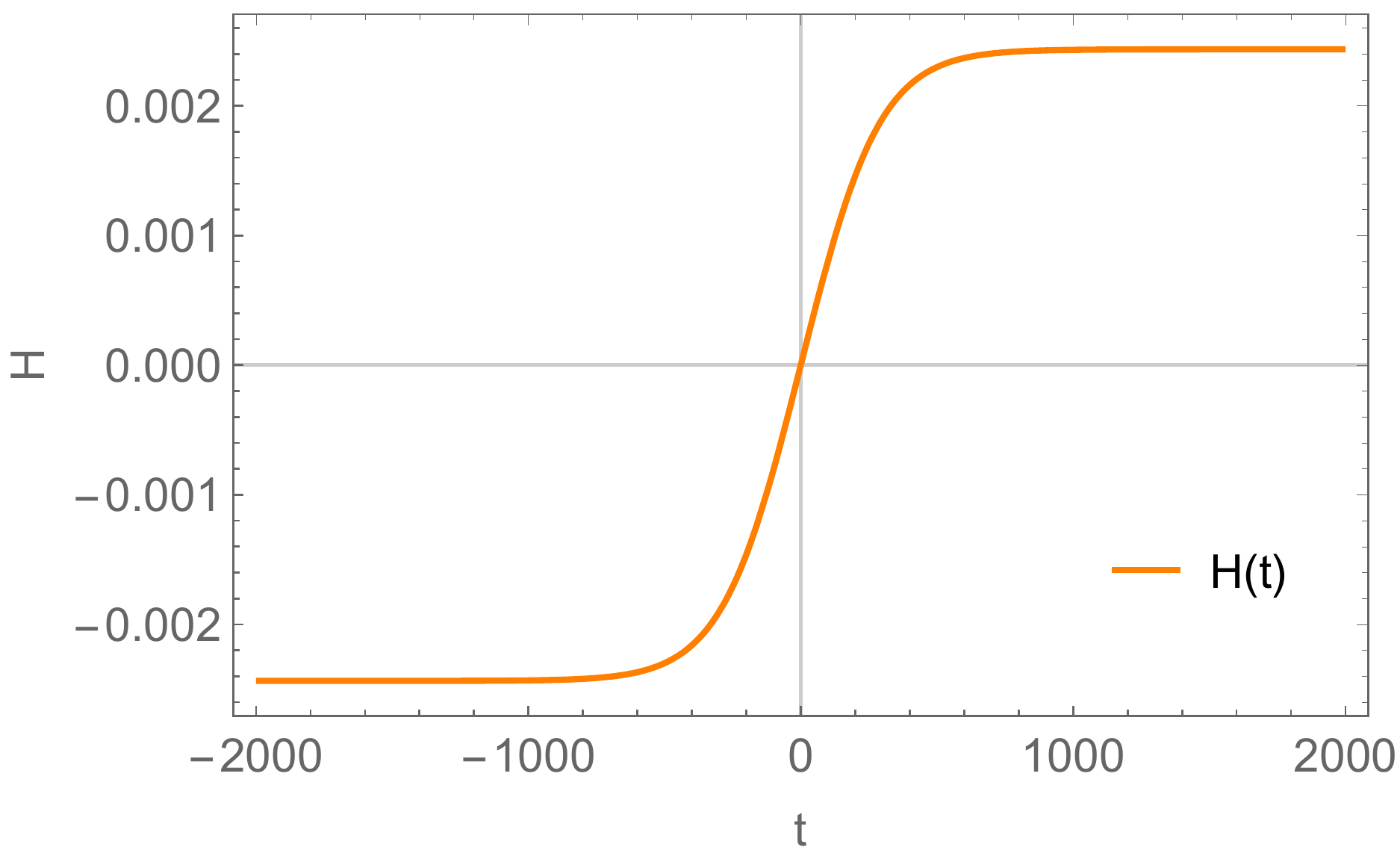}
\hspace{0.5cm}
\includegraphics[height=4.3cm]{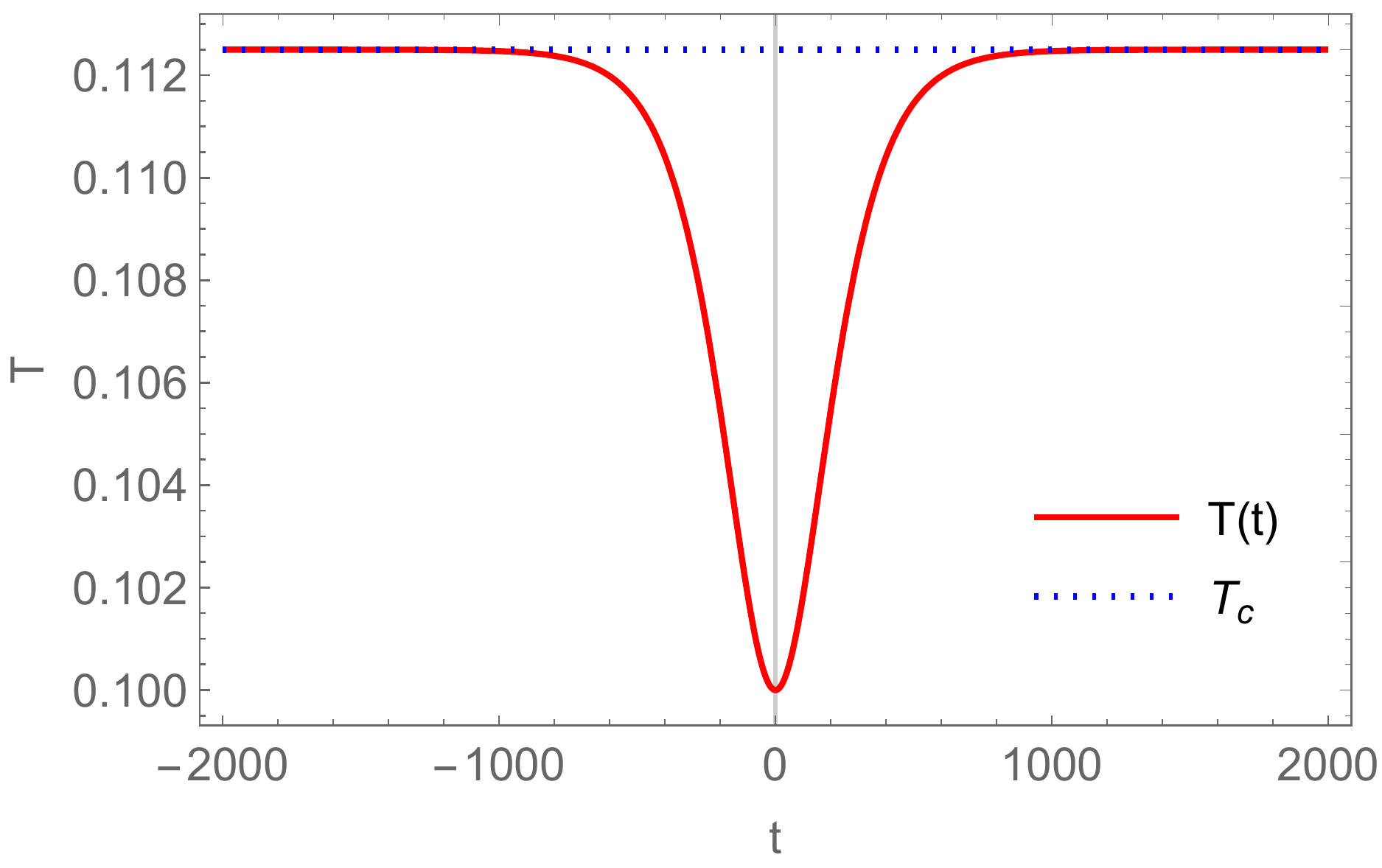}
\caption{\it Numerical solutions of $H(t)$ (left panel) and T(t) (right panel) for $\sigma =  - \delta = 1$ and $B = -0.1 $. {All quantities are expressed in Planck units.} $t = 0$ represents the moment of the bounce. The blue dotted line in the right panel is $T_c$, the maximal allowed temperature that is asymptotically reached in the infinite past and future.} 
\label{fig:HTnumerical}
\end{figure}

\begin{figure}
\centering
\includegraphics[height=6cm]{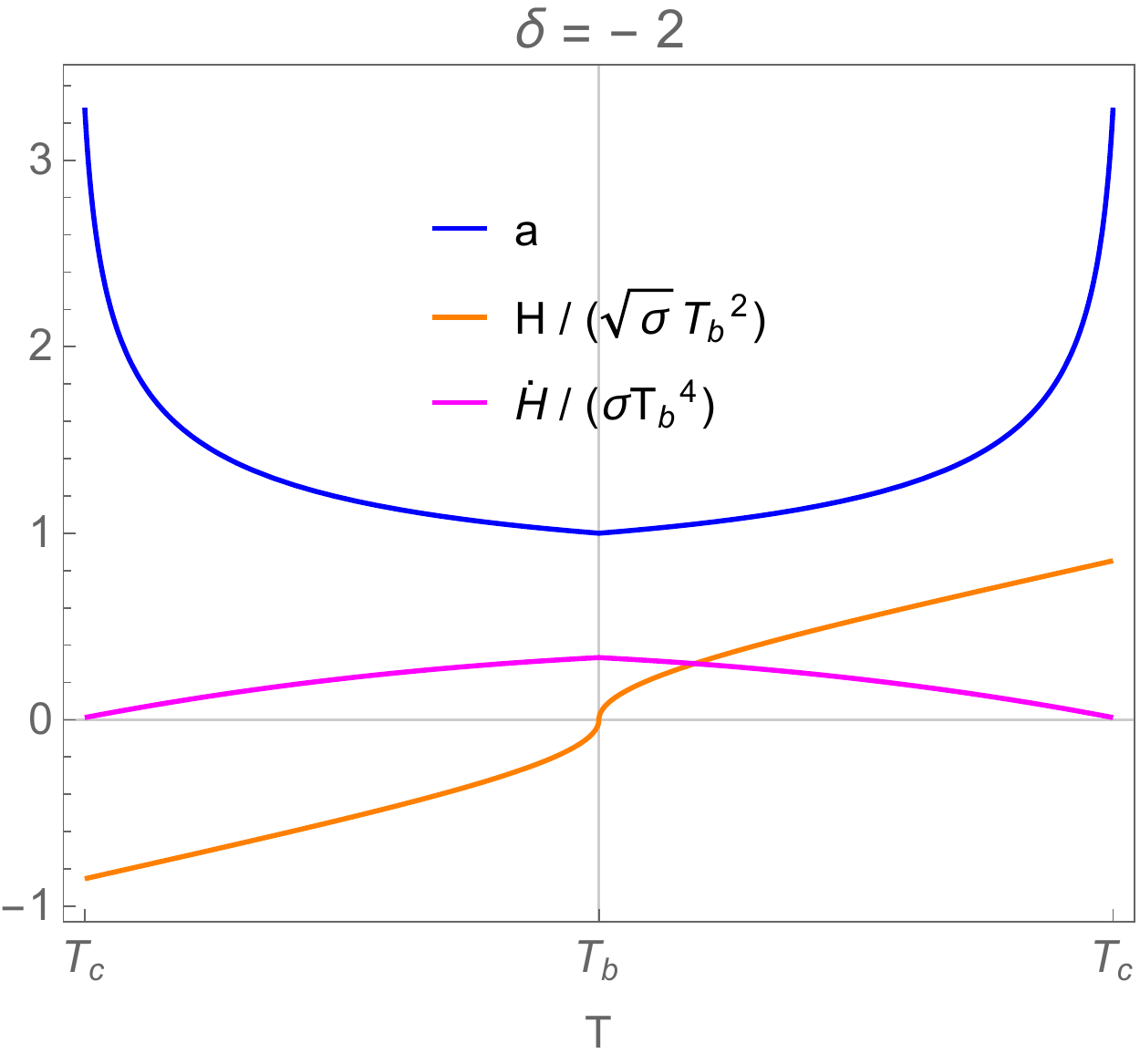}
\caption{\it The scale factor $a$, the normalized Hubble parameter $H$ and its time derivative $\dot{H}$ as a function of temperature in arbitrary units for the $\delta=-2$ case. Time flows from left to right.} 
\label{fig:ahtemp}
\end{figure}

Requiring $H(T_b)=0$ can be satisfied for $\left.\frac{dx}{dt}\right|_{t = t_b} = 0$, or $\delta=-3$.\footnote{$\delta\rightarrow0$ is a smooth limit of the $\frac{dx}{dt}=0$ case.}
 The latter case will be dealt separately. Therefore, the temperature must have an extremum at the bounce. The fact that the temperature is extremal at the bounce is crucial. Without it, there is nothing to limit the energy density from evolving to negative values and being pathological. Because of the extremality of the temperature this never occurs. The temperature dynamically reaches a minimum (or maximum in some cases in the next section) at the bounce and always evolves with strictly positive energy density and positive temperature. One can show, that in fact $T(t)$ has a minimum at $t = t_b$, by noticing that from eq. \eqref{eq:Fried2} for $T \simeq T_b$ one finds
\begin{equation}
\dot{H} \simeq (3+\delta) \ddot{T}(t) > 0 \qquad \Rightarrow \qquad \left.\ddot{T}(t)\right|_{t = t_b} > 0 \, .
\end{equation}
From Eqs. (\ref{eq:rhou},\ref{eq:pu},\ref{eq:Fried2}) one finds
\begin{equation}
\dot{H} = 0 \qquad \Leftrightarrow \qquad x = x_c = \left(\frac{4}{3}\frac{\delta +3}{\delta +4}\right)^{1/\delta } \, . \label{eq:Tcpure}
\end{equation}
Note that $x_c > 1$ for $\delta \in (-3,0]$. Therefore, the temperature grows after the bounce, until it reaches its maximal value $x = x_c$, for which $H = const$. This is also true for the contracting phase of the evolution. 
Integrating the first Friedmann equation, \eqref{eq:Fried1}, we get an analytical expression for time as a function of temperature: 
\begin{equation}
t_{\pm}(T) = \pm\int \sqrt{\frac{3}{\rho(T)}}\frac{a_T}{a}dT \, , \label{eq:tofT}
\end{equation}
where $\rho = \rho_u$, 
\begin{equation}
t_\pm = \pm \frac{\frac{\delta +4}{\delta-2} x^{\delta } F_1\left(\frac{\delta -2}{\delta };\frac{1}{2},1;2-\frac{2}{\delta };x^{\delta },y^{\delta}
\right)+2 F_1\left(-\frac{2}{\delta };\frac{1}{2},1;\frac{\delta -2}{\delta };x^{\delta },y^{\delta}
\right)}{4 \sqrt{\sigma/3} T_b^2 x^2} \, ,
\end{equation}
where $F_1$ is the Appell hypergeometric function. 
Besides the case of $3 + \delta \ll 1$, the Hubble parameter and scale factor are given to a good approximation by:
\begin{eqnarray}
H(t)  =  H_c \tanh\left(\frac{3}{2} H_c \,t\right) \, ,\qquad 
a(t)  =  \cosh\left(\frac{3}{2} H_c \,t\right)^{2/3},
\end{eqnarray}
where
   $H_c = H(t_c) = \frac{1}{3}\sqrt{\frac{-\delta \sigma}{\delta+4}}T_c^2  = \frac{1}{3} \sqrt{\frac{-\delta  \sigma
   }{\delta +4}} \left[\frac{4(\delta+3)}{3(\delta+4)}\left(-\frac{\sigma}{B} \right)\right]^{\frac{2}{\delta}}$.
A plot of the (exact numerical) Hubble parameter $H$ as a function of time is given in the left panel and the temperature as a function of time in the right panel of  Figure \ref{fig:HTnumerical}. An example of the dependence of $a,H,\dot{H}$ on the temperature is given in Figure \ref{fig:ahtemp}.  

Hence, considering the Banks-Zaks theory, slightly removed from its fixed point allows an interesting "dS Bounce" scenario. First, it is not based on a scalar field, but rather the averaged behavior of a non-Abelian gauge theory with a suitable number of fermions \cite{Banks:1981nn}. Second, the epoch preceding the bounce, is not that of slow contraction, but an exponential contraction. Third, the entire evolution is fully determined. The requirement of a regular bounce implies the existence of both the exponential contraction and the ensuing exponential expansion, i.e. Inflation. Finally, the temperature is bounded from above and below, where the minimal temperature is at the bounce and the maximal at the exponential contraction/expansion. To allow for a parameterically wide range of temperatures, one needs to work in the region of $\delta\rightarrow -3$ that we shall consider now separately as it has a simple analytic solution.  

\subsection{Analytical solution for $\delta = -3$}

\begin{figure}
\centering
\includegraphics[height=5cm]{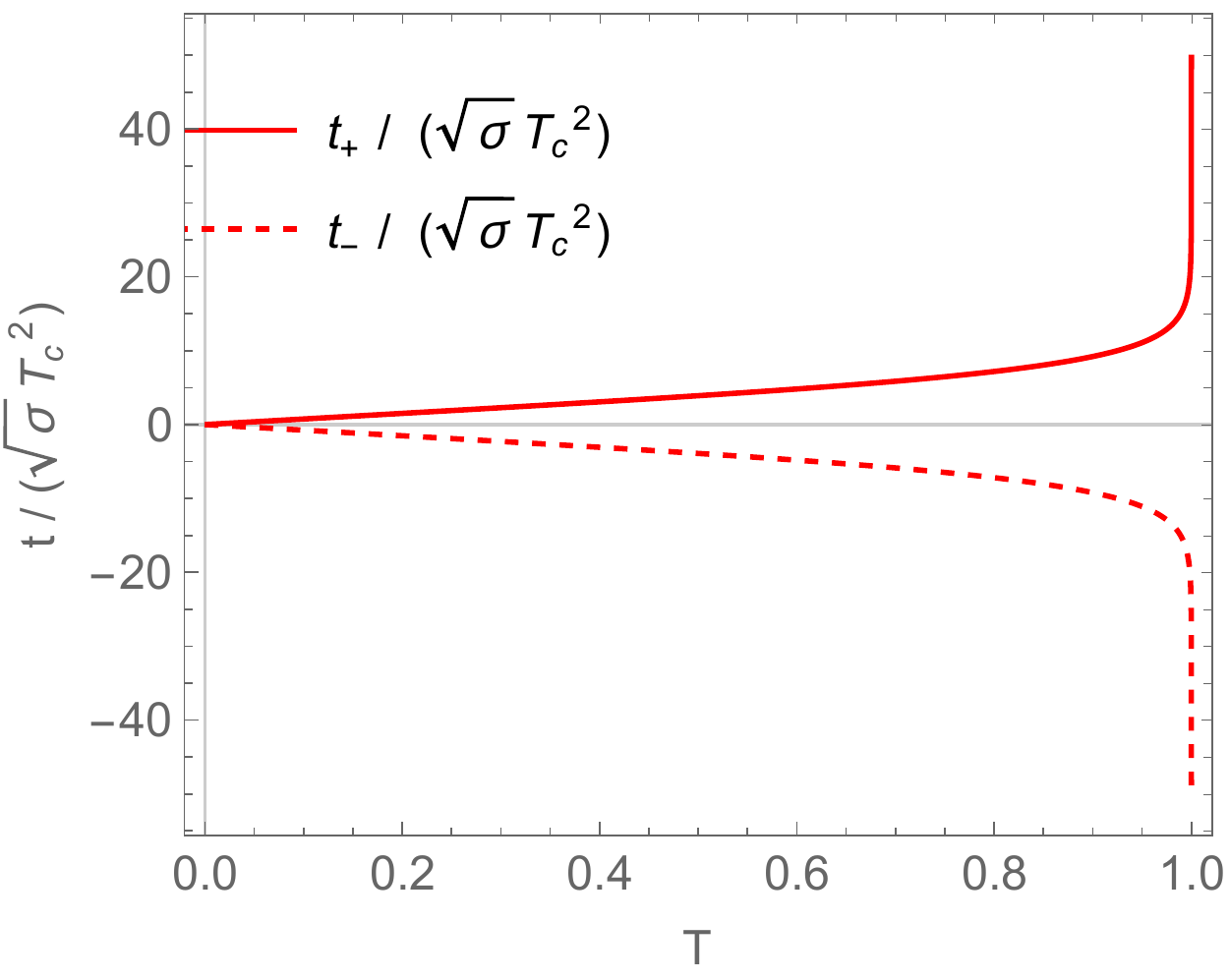}
\hspace{0.3cm}
\includegraphics[height=5cm]{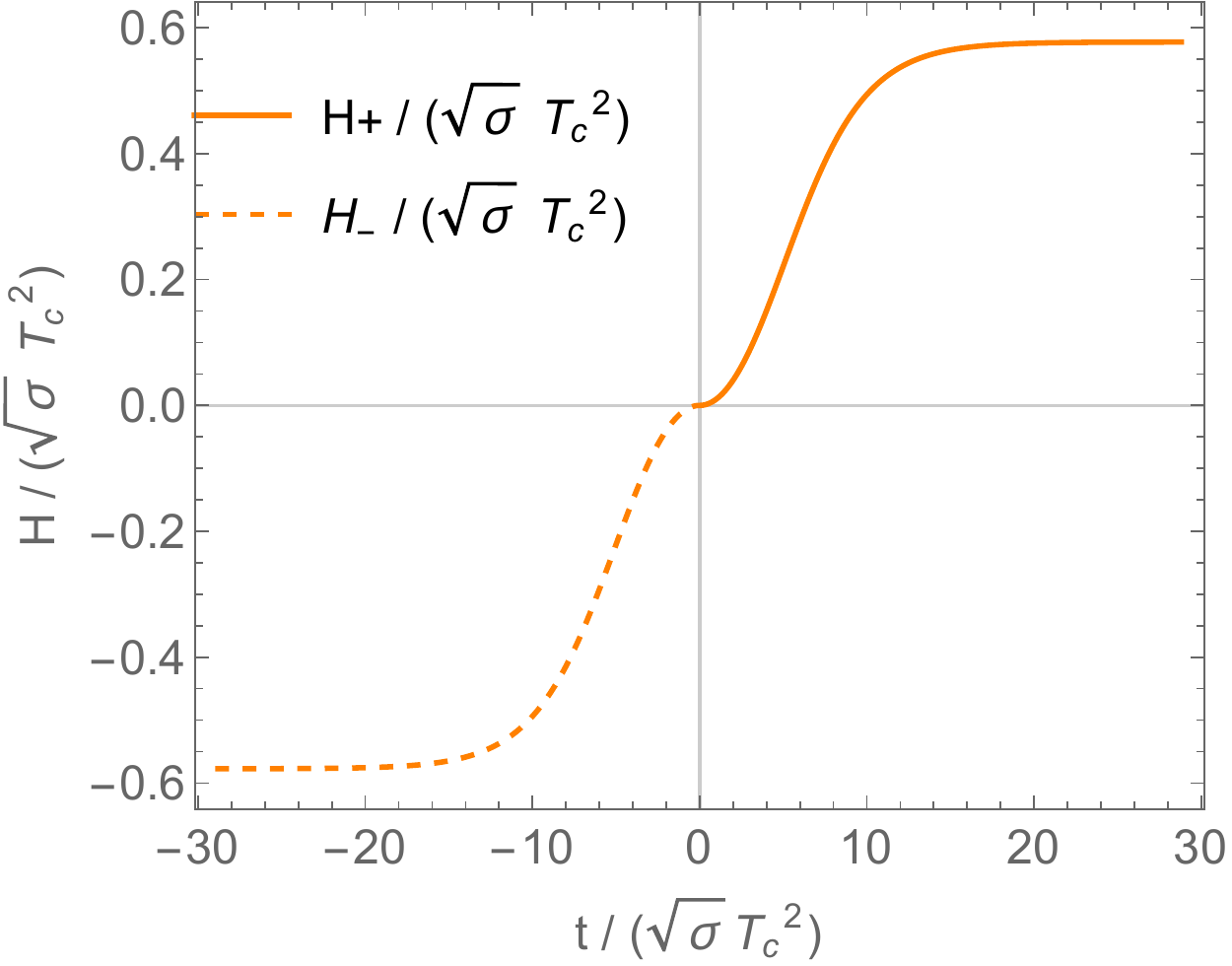}
\caption{\it The analytical solutions for $\delta = -3$. Normalized time vs. normalized temperature in the left panel. In the right panel we have normalized Hubble parameter vs. normalized time.} 
\label{fig:rhoNanalyticaldelta-3}
\end{figure}

The case of $\delta=-3$ is specifically interesting for several reasons. First and foremost, the energy density of unparticles reduces to that of normal radiation. Only the pressure deviates from that of radiation. Second, the temperature at the bounce is reduced to zero. There still is an upper bound on the temperature for the asymptotic de Sitter evolution. Finally, the expressions of the physical quantities are rather simple. As a result, this analytical solution is a limiting simple case of the above analysis. 
For $\delta = -3$, Eq. \eqref{eq:contu} implies 
\begin{equation}
a = \left(1-y^3\right)^{-1/3}  \, ,\qquad H = \frac{y^2}{1 - y^3}\dot{y} \, , \qquad y = \left(1-\frac{1}{a^3}\right)^{1/3} \label{eq:scale-3} \, ,
\end{equation}
where as before $y= T / T_c$ and $T_c = (A/(4 \sigma))^{1/3}$. Note that in the $\delta = -3$ case the bounce appears for $T=T_b = 0$ and therefore $\rho_u(T_b) = p_u(T_b) = 0$, while $\dot{y}$ is not necessarily zero.  
This is in contrast to the previous analysis. 
In fact one obtains discontinuity of $\dot{y}$ at the bounce (but not of $y$ {or $H$}!), since $\dot{y} = \pm \sqrt{\sigma/3}T_c^2 \left(1-y^3\right)$ and therefore $\dot{y}(t \to t_b) \to \pm \sqrt{\sigma/3}T_c^2$. {Nevertheless, one can still start from the contracting Universe and smoothly evolve the system through the bounce, towards the expanding solution. Therefore, this discontinuity is not a problem in any way.} The $\delta = -3$ case is the only one, for which one may obtain a bounce with $T_b = 0$. In all other cases $T=0$ is indeed a solution of $\rho_u = 0$, but it cannot lead to a physical bounce, since it has negative energy density $\rho_u(T) < 0$ for $0<T<T_b$.
From \eqref{eq:tofT} one finds
\be
t_\pm = \pm \frac{1}{2\sqrt{3\sigma}T_c^2}\left(\ln \frac{1-y^3}{(1-y)^3} +2\sqrt{3}\tan ^{-1}\left(\frac{2}{\sqrt{3} }(1+y)\right)-2\pi\right) \, ,
\ee
where the constant part was added to synchronize time and temperature, $t=0$ with $T=0$. Time as a function of temperature and the Hubble parameter as a function of time  for the case of $\delta=-3$ are plotted in Fig. \ref{fig:rhoNanalyticaldelta-3}.

To summarize, we get the following solution. An exponential contracting phase at $T=T_c$ with constant $H$, followed by a bounce at $T=T_b$ which rapidly evolves into an inflationary phase with the same magnitude of $H$ but with a positive sign, again with $T=T_c$. The asymptotic future is de Sitter space and the asymptotic past is exponential contraction, (not Anti de Sitter), a "dS Bounce".

This dS Bounce has several useful features. First, the energy density $\rho$ is always positive. The temperature is always positive but bounded throughout the evolution $T_b\leq T \leq T_c$, and correspondingly $x\geq1$ and $y\leq1$. However, except the limiting case of $\delta\rightarrow -3$, we have $T_c\sim \mathcal{O}(1) T_b$. Another peculiarity is that the minimal temperature is at the bounce. The reason is the peculiar form of energy. Rewriting the NEC, we get
\be
\rho_u+p_u=\frac{4\sigma}{3}T^4\left(1-y^{\delta}\right)<0
\ee
Since in our scenario $T_b/T_c<y<1$ and $\delta<0$, the expression in the brackets is always negative, so throughout the evolution the NEC is violated, saturating it only at $T=T_c$. This is depicted in right panel of Fig. \ref{fig:Toft}. Regarding the $\delta=-3$ case, we want to emphasize that this unique solution (i.e. the bounce with both kinetic and potential energy densities equal zero) is possible only due to the fact that NEC is violated and therefore the energy density of the Universe may grow together with the scale factor.

\subsection{Approximate solutions}

Around the bounce and the de Sitter phase one can obtain a more tractable solution for the time-temperature relation than Appell functions. This will be useful for calculating perturbations. In particular we can write a simple expression for $T(t)$. Let us note that for $T \simeq T_b$ one finds $\dot{T} \simeq 0$. In such a case, from Eqs. \eqref{eq:Fried2} and \eqref{eq:scale} one finds
\begin{equation}
\ddot{T} \simeq -\frac{\delta \, \sigma \, T_b^5}{6 (\delta +3)^2} \, ,\label{eq:ddT}
\end{equation}
which gives the following approximate temperature, Hubble parameter and scale factor as a function of time:\footnote{ {Note that possible divergences due to $\delta\rightarrow -3$ in the above equations are not true divergences. In the limit of $\delta= -3$, $T_b=0$ and one should rederive the approximate formulae using $A$ instead of $B$ in equations (\ref{eq:rhou},\ref{eq:pu}).}} 
\begin{eqnarray}
T &\simeq& T_b\left(1-\frac{ \delta \, \sigma  \,T_b^4}{12 (\delta +3)^2} t^2 \right)=T_b\left(1+\frac{H_c^2}{\delta+3}\left(\frac{4(\delta+3)}{3(\delta+4)}\right)^{-1-\frac{4}{\delta}}t^2\right)  \, , \label{eq:analyticalToft}\\ 
H(t) & \simeq & -\frac{\delta  \sigma  T_b^4}{6 (\delta +3)}t =2H_c^2\left(\frac{4(\delta+3)}{3(\delta+4)}\right)^{-1-\frac{4}{\delta}}t\, ,\label{eq:Happrox}\\
a(t) &\simeq & 1-\frac{\delta  \sigma  T_b^4}{12 (\delta +3)} t^2 = 1+H_c^2\left(\frac{4(\delta+3)}{3(\delta+4)}\right)^{-1-\frac{4}{\delta}}t^2 \, . \label{eq:aapprox}
\end{eqnarray}

{Let us stress that despite the relatively fast changes in temperature around the bounce, that can be seen from the above figures and formulae, one still obtains the adiabatic evolution of the temperature of unparticles, since the condition $T\sim\omega \gg H$ is satisfied for all $M_p>T>T_b$.}

\begin{figure}
\centering
\includegraphics[height=4.6cm]{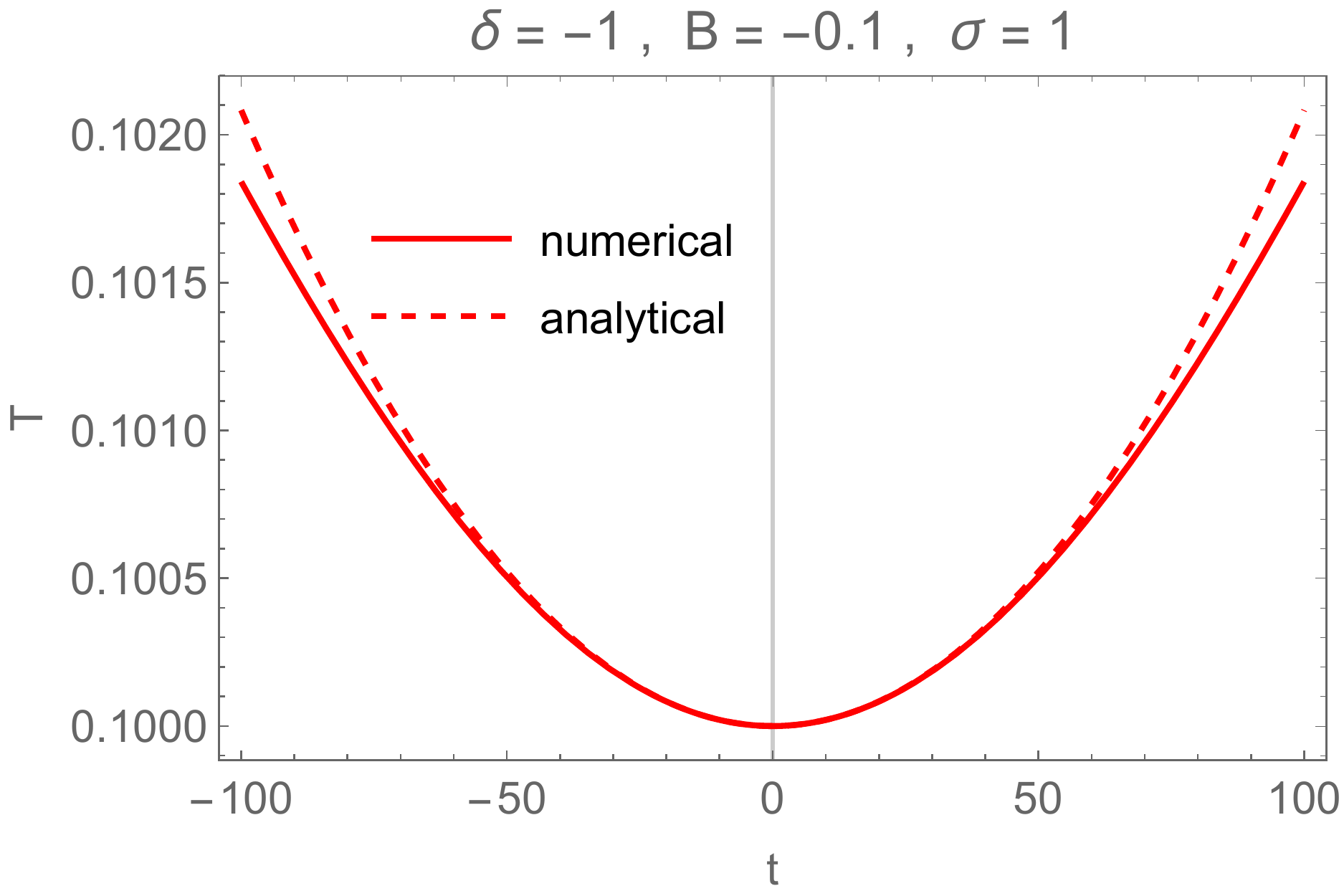}
\hspace{0.5cm}
\includegraphics[height=4.6cm]{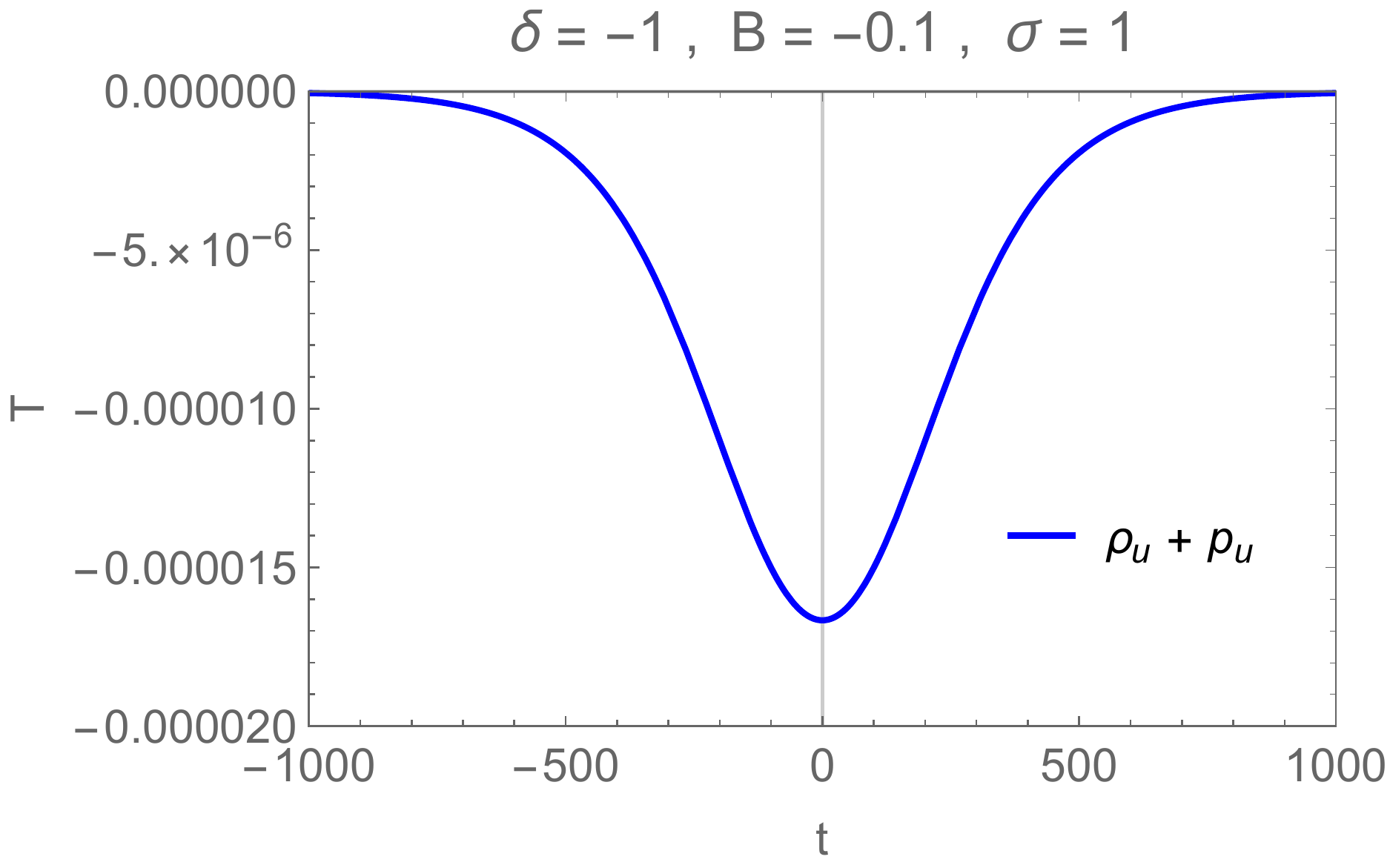}
\caption{\it Left panel: Numerical solutions of $T(t)$ for $\sigma= - \delta = 1,\,B=-0.1$ together with the analytic approximation from Eq. \eqref{eq:analyticalToft} (solid and dashed lines respectively). {All quantities are expressed in Planck units.} $t = 0$ represents the moment of the bounce. Right panel: $\rho_u + p_u$  for unparticles for $\sigma = -\delta = 1,\, B=-0.1$. Note $\rho_u  + p_u< 0$ always, which violates the NEC. } 
\label{fig:Toft}
\end{figure}

The numerical result of $T(t)$ as well as the analytical approximation has been presented in the left panel of  Fig. \ref{fig:Toft}.

Similarly, we obtain the asymptotic approximation around the dS phase. To obtain an analytic solution for $t(T)$ one solves the continuity equation (\ref{eq:contu}) in vicinity of $ T_c $. $t(T)$ can be then inverted to get $T(t)$ as follows
\begin{equation}
T(t) \simeq T_c\left(1 - 4^{-\frac{1}{\delta}}a(t)^{-3}\right). \label{eq:TapproxdS}
\end{equation}
It is then straightforward to obtain expression for $H(t)$ as 
\begin{equation}
H(t) \simeq H_c\left( 1 - 2^{1-\frac{2}{\delta}}\left(4+\delta\right)a(t)^{-3}\right), \label{eq:HapproxdS}
\end{equation}
where $a(t) \simeq e^{{H_c}t}$ is the scale factor. 

Since unparticles generate de Sitter-like evolution in the vicinity of $T_c$, it is worth comparing this model with cosmic Inflation. In most inflationary models one defines slow roll parameters $\epsilon = -\frac{\dot{H}}{H^2} $ and $ \eta = -\frac{\ddot{H}}{2\dot{H}H}$ and demands that both $ \epsilon $  and $ |\eta| $ $ \ll 1$. In the case of unparticles the slow roll parameters are given by 
\begin{eqnarray}
 \epsilon & \simeq & -3\times 2^{1-\frac{2}{\delta}}\left(4 + \delta \right)a(t)^{-3}, \\
 \eta & \simeq &-\frac{3}{2}\left(1- 2^{1-\frac{2}{\delta}}\left(4 + \delta \right)a(t)^{-3}\right).
\end{eqnarray} 
One can see that $\epsilon$ is exponentially suppressed, while $\eta  \simeq -\frac{3}{2}$, which makes this case similar to the constant-roll inflation \cite{Motohashi:2014ppa,Yi:2017mxs}. Note that some version of unparticles have already been analyzed in the context of cosmic Inflation \cite{Collins:2008ny}. It is was shown that unparticles by themselves cannot generate Inflation consistent with observational data.

\subsection{Non-bouncing solutions for unparticles} \label{nonbounce}

To complete the "phase diagram", let us briefly investigate non-bouncing scenarios for the evolution of the Universe. {The $\rho=0$ initial condition can be also obtained for $B<0$ and $T=T_b$, which for $\delta \notin [-3,0]$ gives a recollapse with the following cosmic scenarios \footnote{{In such a case one finds no bounce and $T=T_b$ represents a recollapse.}} 
\begin{itemize}

\item    
     For $B<0,\,\delta < -3$ the temperature {obtains its minimum at the recollapse and grows while the Universe contracts. Thus, shortly after the recollapse one can consider the $x \gg 1$ limit, which gives}
    \begin{equation}
        T \simeq \left(1-2 \sigma \,  t \, T_b^2 \right)^{-1/2} \, .
    \end{equation}
    This solution has a pole, which means that the temperature reaches infinity at finite time and one reaches a curvature singularity. This case is presented in Fig. \ref{fig:alpha=0} in red. 
    
    \item For $B<0,\,\delta > 0$ one obtains a recollapse at $T=T_b$ with a maximum of $T$ at the recollapse. While the Universe contracts, the temperature drops to $T_p= \left(-\frac{4 \sigma}{B (\delta +4) }\right)^{1/\delta }=\left(\frac{4}{4+\delta}\right)^{1/\delta}T_b<T_b$ for which $a_T=0$ and both $T(t)$ and $H(t)$ become discontinuous. Hence, this solution is unphysical. This scenario is presented in Fig. \ref{fig:alpha=0} in brown. 
\end{itemize}

{For $B>0$ one cannot obtain $\rho = 0$ for $T\neq 0$ and therefore one cannot obtain a bounce or a recollapse. Depending on the value of $\delta$ one finds the following scenarios for the evolution of the Universe:}

\begin{itemize}
    \item For $B>0, \delta \geq-3$ one finds $a_T <0$ for all $T$, so $a\to 0$ as $T \to \infty$. 
    This is simply a Big Bang Universe, i.e. going backwards in time from today we shall reach an infinitely hot dense Universe with a curvature singularity. $T$ is unbounded, as can be seen since $T_c$ is complex or negative. The Big Bang scenario is presented in Fig. \ref{fig:alpha=0} in white. 
    \item For $B>0, \,\delta < -4$ one obtains a Big Bang scenario in the $T\to \infty$ limit. Nevertheless, for $T \to T_p$ one finds $a_T \to 0$ which again leads to unphysical discontinuity of $T$ and $H$. This case is presented in Fig. \ref{fig:alpha=0} in blue. {One can avoid the discontinuity by assuming that $T<T_p$ throughout the entire evolution of the Universe. In such a case $T\to0$ and $\rho \to \infty$ while $t \to \infty$ and $a\to \infty$.} 
   \item {For $B>0, \, \delta \in \left[ -4, -3\right)$ one obtains real $T_c$ but $T_b$ is complex. Therefore the de Sitter phase can be reached, but one cannot obtain a bouncing scenario. This is the pink region in Fig. \ref{fig:alpha=0}. In such a case the evolution of the Universe depends on initial value of the temperature denoted as $T_i$:}
   \begin{itemize}
 \item {For $T_i > T_c$ the temperature decreases towards $T_c$ and one obtains constant-roll de Sitter expansion with $\epsilon>0$.} {In the $t \to -\infty$ limit one finds $T\to \infty$ and therefore the Universe starts from the Big Bang singularity. The NEC is never violated throughout all of the evolution. Of particular phenomenological interest is the $\delta=-4$ case that corresponds to a universe filled with radiation and cosmological constant. One finds $a\propto 1/T$ and $T$ decreases throughout the evolution of the Universe. The constant is a potential candidate for Inflation or late-time acceleration.}
 \item {For $T_i < T_c$ the temperature increases while the Universe grows. For $T\lesssim T_c$ one again obtains constant-roll de Sitter expansion with $\epsilon<0$. In this case the late time evolution is the same as in Eqs.~(\ref{eq:TapproxdS}\,,\,\ref{eq:HapproxdS}).} {Since a real $T_b$ does not exist, one does not obtain any lower bound on $T$. In fact for $t \to -\infty$ one finds $T \to 0$ and $\rho_u \to 0$. In such a case one starts from the empty Minkowski Universe and increases the temperature up to the $T\lesssim T_c$ limit. The NEC is always violated throughout all of the evolution.}
   \end{itemize}
\end{itemize}

\begin{figure}[H]
\centering
\includegraphics[height=5cm]{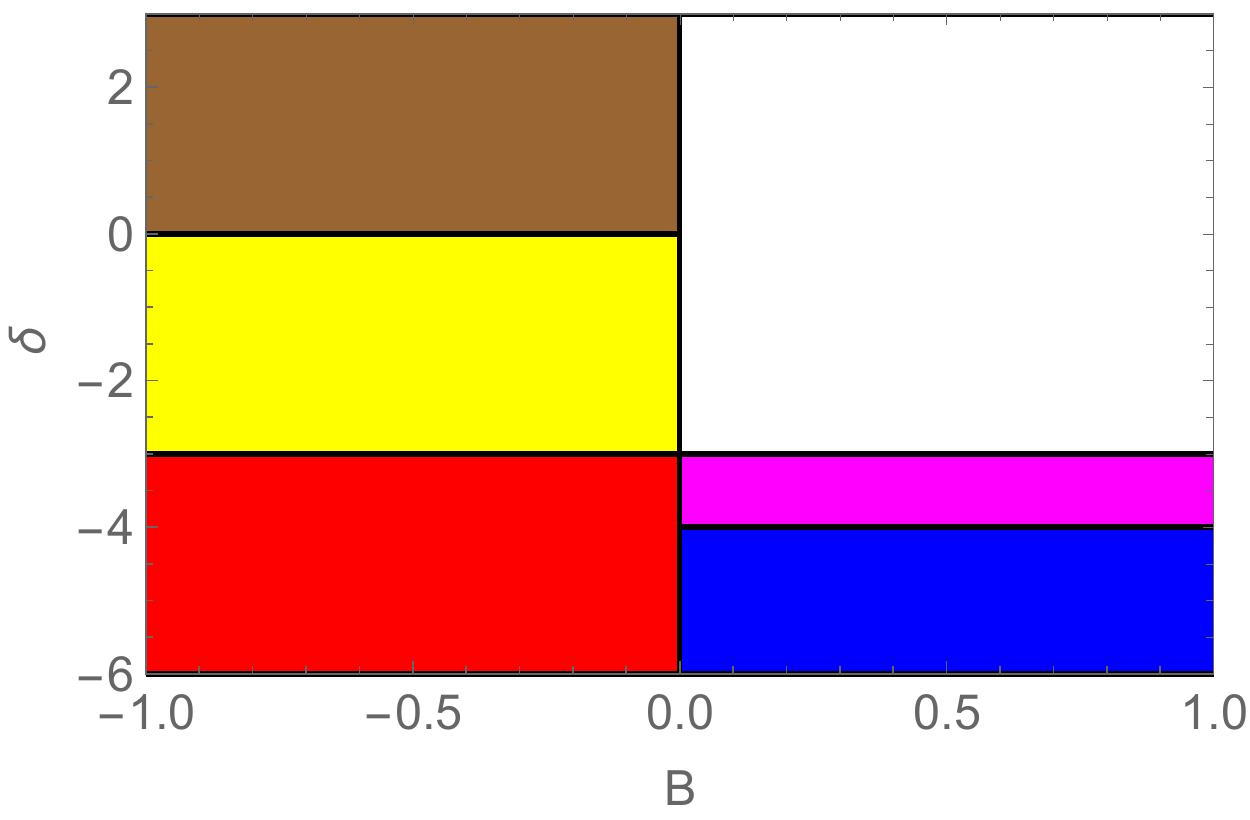}
\caption{\it Different cosmological scenarios for the Universe filled with unparticles. White color labels the Big Bang Universe with the singularity in the past. Red and brown represent a (re)collapsing unstable Universe. The blue color is a Universe with a Big Bang and an instability of $T$, while the yellow color represents the case we have thoroughly discussed of exponential contraction, bounce and de Sitter expansion, i.e. the dS bounce. The pink region corresponds to future asymptotic dS phase that was preceded by either a Big Bang or an empty Minkowski Universe, depending on initial conditions.} 
\label{fig:alpha=0}
\end{figure}

To conclude, we have analyzed the possible evolution of a universe filled with unparticles. It results either in a dS Bounce, a standard Big Bang model (either asymptoting to dS or not), or pathological cases where the Hubble parameter and temperature are discontinuous. The non-singular solutions resulted in a temperature bounded from above and below as one may expect in order to have a finite energy density at all times. However, the allowed range of temperatures is parameterically large only for $\delta \simeq -3$.


\section{Universe evolution with unparticles and a perfect fluid} 
\label{sec:fluid}
It is realistic to assume that besides unparticles, the Universe was also filled with additional fields, which could be described as a perfect fluid with $\rho_f(t) = \rho_{f0} \left(\frac{a}{a_0}\right)^{-3(1+w)}$ and $p_f = w\, \rho_f$, where $\rho_{f0}\,,\,a_0$ are the values of the energy density and scale factor at some time $t=t_0$ and $w$ is the equation of state parameter of the perfect fluid. For simplicity we limit ourselves to $-1\leq w \leq 1$. The Friedmann equations (\ref{eq:Fried1}\,,\,\ref{eq:Fried2}) are now with $\rho = \rho_u+\rho_f$ and $p = p_u+p_f$. {We still assume that $T<\Lambda_\mathcal{U}$ and hence unparticles and the perfect fluid are decoupled.}

As in Sec. \ref{sec:uonly}, we want to obtain a bouncing scenario generated by unparticles, which requires $\rho_b = 0$ and $p_b < 0$. Defining $\alpha \equiv \rho_{f0}/(\sigma T_0^4)$, $\alpha$ gauges the amount of the perfect fluid with respect to the (standard radiation part) of the unparticles at $t=t_0$. As such it gauges the importance of unparticles at that time $t_0$, whether they are the dominant contribution to the pressure and energy density or some negligible contribution. Our analysis shows that the existence of an additional perfect fluid, results in additional solutions beyond the single bounce. In all our solutions we demand the positivity of the total energy density $\rho\geq 0$ and the temperature $T\geq 0$. Let us assume that $t=t_0$ is a moment of a regular bounce or recollapse of the universe, i.e. $H=0$. In such a case one requires $\rho(t_0)=\rho_0 = 0$, which implies that the temperature $T_0=T(t_0)$ will be 

\begin{equation}
T_0 = \left( -\frac{\sigma}{B}(1+\alpha)\right)^{\frac{1}{\delta}} \, .
\end{equation}

Note that for $\alpha = 0$ one finds $T_b=T_0$ reproducing the unparticles only case. In order to simplify further calculations let us define again $x \equiv \ T/T_0$. As in the previous section, one can obtain $a = a(x)$ from the continuity equation of unparticles:\footnote{We want to emphasize that throughout the paper $T$ is the temperature of unparticles, but not necessarily the temperature of other fluids. In the presence of multiple decoupled fluids as we have here, one expects each of them to have a different temperature. A simple example is radiation, where one obtains $a\propto 1/T_R$, where $T_R$ is the temperature of radiation, which is obviously different from the temperature of unparticles.} 

 \begin{eqnarray}
a(x) =\frac{1}{x} \left(\frac{3 \alpha  (\delta +4)-\delta}{3(\delta +4)(1+\alpha) x^{\delta }-4 (\delta +3)}\right)^{1/3} \label{eq:scT}, \\
H(x) = -\frac{(\delta +3) \left((\delta +4)\left(1+\alpha \right) x^{\delta}-4\right)}{x \left(3 (\delta +4)\left(1+\alpha\right)
   x^{\delta }-4 (\delta +3) \right)} \frac{dx}{dt}. \label{HT}
  \end{eqnarray}

The $t(T)$ is derived again from \eqref{eq:tofT}. The normalization of $a(x)$ has been chosen to give $a(x=1) = 1$. We then express the energy density and pressure as function of $x$:
\begin{eqnarray}
 \rho &=&\sigma  T_0^4 x^4 \left(\alpha  x^{3 w-1} \left(\frac{4 (\delta +3)-3 (\alpha +1) (\delta +4) x^{\delta }}{\delta -3 \alpha  (\delta +4)}\right)^{w+1}\!-(\alpha +1) x^{\delta }+1\right),\label{eq:rhoT}\\
p &=& \sigma  T_0^4 x^4 \left(\alpha  \, w \,  x^{3 w-1} \left(\frac{4 (\delta +3)-3 (\alpha +1) (\delta +4) x^{\delta }}{\delta -3 \alpha  (\delta +4)}\right)^{w+1}\!\!-\frac{(\alpha +1) x^{\delta }}{\delta +3}+\frac{1}{3}\right).\cr 
\end{eqnarray}

Obviously $x=0$ and $x=1$ are solutions  to $\rho=H=0$, but there may be other solutions.
From here we proceed as follows: 

\begin{enumerate}
    \item We consider the $x=1$ solution, as a starting point and calculate the pressure at this point. 
     \begin{equation}
p_0 = p(T_0) = \sigma T_0^4 \frac{ \delta +3 \alpha  [(\delta +3) w-1]}{3 (\delta +3)} \, .
\end{equation}
A negative pressure will correspond to a bounce (i.e. the Universe will initially expand after deviating from $x=1$) and positive pressure to a recollapse (i.e. the universe will initially contract following $x=1$).
We get the following condition on $w$:
\be
p_0<0\Leftrightarrow w<\frac{1}{\delta+3}\left(1-\frac{\delta}{3\alpha}\right)
\ee
From \eqref{HT} we have seen that
 $H(x=1)=0$ corresponds to an extremum of the normalized temperature $\frac{dx}{dt}=0$.\footnote{Other solutions are $\delta = -3$ or $\delta = -\frac{4\alpha}{4+\alpha}$. The case of $\delta = -3$ will bounce only with vanishing temperature at the bounce as in the unparticles only case. For a finite scale factor at the bounce, this will happen only for $\rho_{f0} = 0$ at the bounce, which implies $\rho_f\equiv 0$, so we are back with the unparticles only case.  
 The case $\delta = -\frac{4\alpha}{1+\alpha}$ is analyzed separately in subsection \ref{sec:special}.} 
Denoting \eqref{HT} as $H\equiv f(x)\dot{x}$ we have at $x=1$:
\be
-\frac{1}{2}p_0=\dot{H}_0=f'_0\dot{x}_0^2+f_0\ddot{x}_0=f_0\ddot{x}_0
\ee
Hence, one obtains a Bounce or a recollapse with a minimum or maximum of the temperature depending on the parameters in $f_0$.

\item We numerically integrate the equations from $x=1$.

The bouncing solutions $(\rho_0=0,\, p_0<0)$ are always stable. They lead to a cyclic scenario, a dS Bounce, as in the unparticles only case, or a single "normal bounce" 
where after $H$ reaches some maximum, it gradually decreases as in the Hot Big Bang scenario. 
The recollapsing solutions  $(\rho_0=0,\, p_0>0)$ are either cyclic, reach a curvature singularity at finite time or are unstable.

\item We then checked other roots of $\rho=H=0$ and got qualitatively the same scenarios. So for a given set of parameters one can have several possible branches of universe evolution depending on the initial temperature. 
\end{enumerate}

Table \ref{table1} lists all scenarios as a function of $\alpha,\delta,w$ assuming the initial condition of $x=1$. An example of the different scenarios for $\alpha=1$ is presented in Fig. \ref{fig:alpha1}: Green corresponds to a bouncing scenario with maximal temperature at the bounce, yellow to a bounce with a minimal temperature at the bounce, red to a recollapse with a mininal temperature at the recollpase, and brown to a recollapse with maximal temperature at that point. 
The yellow and brown cases are somewhat counter-intuitive, since they indicate that $T$ grows while $H>0$ and $T$ decreases for $H < 0$. Note that this counter-intuitive behavior also appears for the bounce generated only by unparticles described in the previous section. 
\begin{center}
\begin{table}[H]
\caption{Conditions for cosmological scenarios for different values of parameters for Unparticles + fluid. The colors, in accord with Figure \ref{fig:alpha1}, label the physical evolution of the Universe at $T=T_0$: Green - bounce at maximal temperature, yellow - bounce at minimal temperature, red - recollapse at minimal temperature, and brown - recollapse at maximal temperature.} 
\vspace{ 1 cm}
\begin{tabular}{|l|l|l|l|l|}
\hline
\bf{\hspace{2em}$w$} & \bf{\hspace{1em}$ \alpha$} & \bf{\hspace{2em}$\delta$} & \bf{Color/Cosmology} \\ \hline
\multirow{11}{*}{$\left( -1\,,\, \frac{3\alpha -\delta}{3\alpha \left(\delta + 3\right)}\right)$} & \multirow{2}{*}{$\left(0\,,\, \frac{1}{3}\right]$} & \multirow{2}{*}{$\left( -\frac{4 \alpha}{1 + \alpha}, - \frac{12 \alpha}{3\alpha -1} \right)$} & \multirow{5}{*}{Green, Bounce @ max T} \\
 &  &  &  \\ \cline{2-3}
 & \multirow{2}{*}{$\left[\frac{1}{3}\,,\, 3 \right]$} & \multirow{2}{*}{$ \left(- \frac{4\alpha}{1 + \alpha} \,,\, \infty\right)$} &  \\
 &  &  &  \\ \cline{2-3}
 & $\left(3\,,\, \infty \right)$ & $ \left( -3 \,,\, \infty \right)$ &  \\ \cline{2-4} 
 & $\left(0\,,\, \frac{1}{3}\right)$ & $ \left( -3 \, ,\,-\frac{4 \alpha}{ \alpha + 1}  \right)$ & \multirow{6}{*}{Yellow, Bounce @ min T} \\ \cline{2-3}
 & \multirow{3}{*}{$\left[\frac{1}{3}\,,\, 3 \right]$} & \multirow{2}{*}{$ \left(-\frac{12 \alpha}{ 3\alpha - 1} \,,\,  \infty \right)$} &  \\
 &  &  &  \\ \cline{3-3}
 &  & $ \left( -3 \, ,\,-\frac{4 \alpha}{ \alpha + 1}  \right)$ &  \\ \cline{2-3}
 & \multirow{2}{*}{$\left[3\,,\, \infty \right]$} & \multirow{2}{*}{$ \left(-\frac{12 \alpha}{ 3\alpha - 1} \,,\,  \infty \right)$} &  \\
 &  &  &  \\ \hline
\multirow{13}{*}{$\left(\frac{3\alpha -\delta}{3\alpha \left(\delta + 3\right)} \,,\, 1\right)$} & \multirow{2}{*}{$\left(0\,,\, \frac{1}{3}\right)$} & $(-\infty\,,\, -3)$ & \multirow{7}{*}{Red, Recoll. @ min T} \\ \cline{3-3}
 &  & $ \left(-\frac{4\alpha}{1 + \alpha}\,,\, -\frac{12\alpha}{3\alpha -1}\right)$ &  \\ \cline{2-3}
 & \multirow{2}{*}{$\left[\frac{1}{3}\,,\, 3 \right]$} & $ \left( -\frac{12\alpha}{3\alpha -1}\,,\, -3\right)$ &  \\ \cline{3-3}
 &  & $ \left(-\frac{6 \alpha}{ 3\alpha + 1} \,,\,  \infty \right)$ &  \\ \cline{2-3}
 & \multirow{4}{*}{$\left[3\,,\, \infty \right]$} & \multirow{3}{*}{$ \left( -\frac{12\alpha}{3\alpha -1}\,,\, -\frac{4\alpha}{1 + \alpha}\right)$} &  \\
 &  &  &    \\ \cline{3-3} 
 &  & $ \left( -\infty \,,\, -\frac{6\alpha}{1 + 3\alpha} \right)$ & \multirow{6}{*}{Brown, Recoll. @ max T} \\ \cline{2-4}
 & \multirow{2}{*}{$\left(0\,,\, \frac{1}{3}\right]$} & $ \left( -\frac{6\alpha}{1 + 3\alpha}\,,\, -\frac{4\alpha}{1 + \alpha}\right)$ &  \\ \cline{3-3}
 &  & $ \left( -\frac{12\alpha}{3\alpha -1}\,,\,\infty\right)$ &  \\ \cline{2-3}
 & $\left[\frac{1}{3}\,,\, 3 \right]$ & $ \left( -\infty\,,\, -\frac{12\alpha}{1 + 3\alpha}\right)$ &  \\ \cline{2-3}
 & \multirow{2}{*}{$\left[3\,,\, \infty \right]$} & $ \left( -\infty\,,\, -\frac{12\alpha}{1 + 3\alpha}\right)$ &  \\ \cline{3-3}
 &  & $  \left( -\frac{4\alpha}{1 + \alpha}\,,\, -3\right)$ &  \\ \hline
\end{tabular} \label{table1}
\end{table}
\end{center}
\begin{figure}[H]
\centering
\includegraphics[height=7cm]{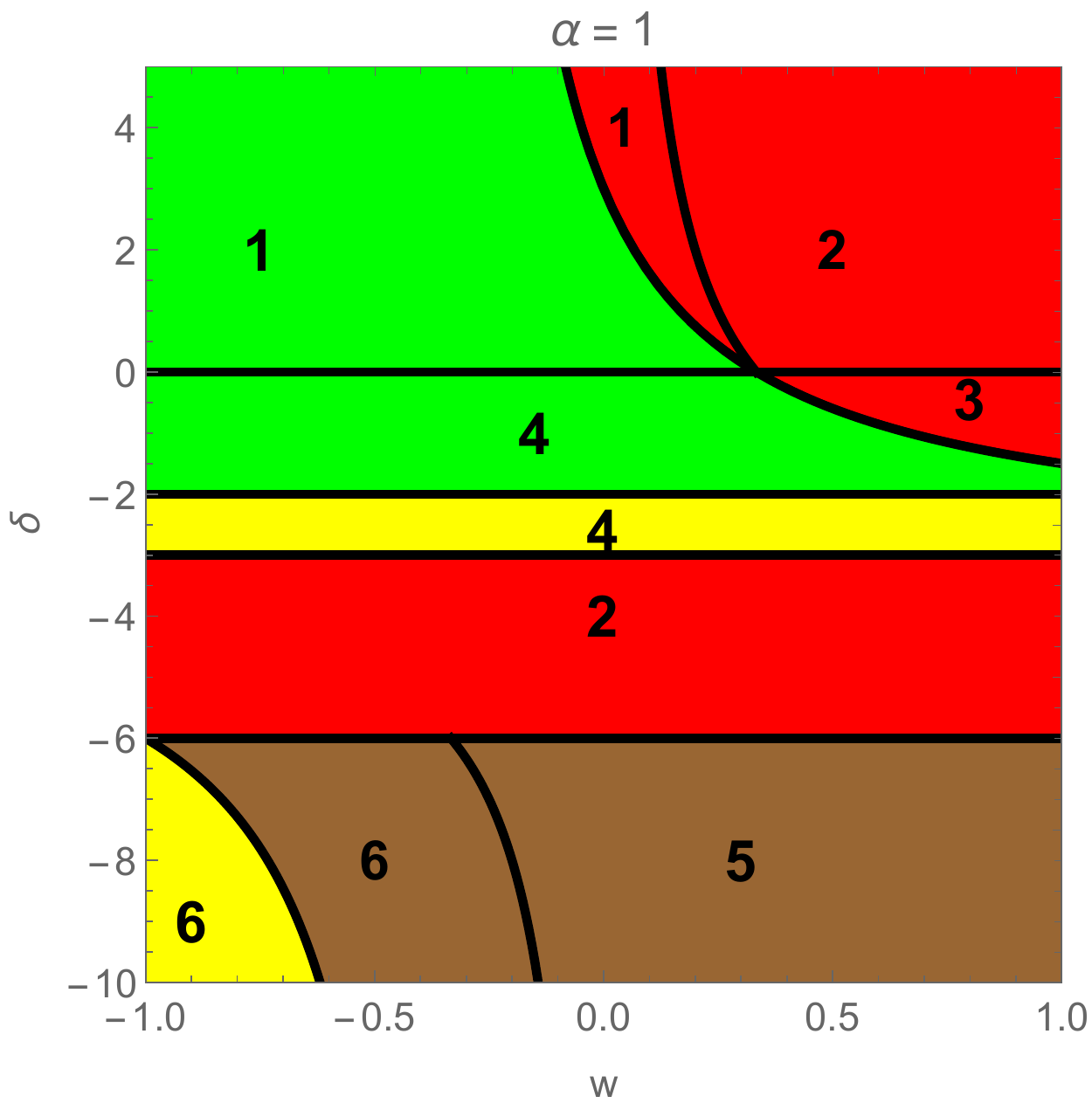}
\caption{\it Classification of different scenarios at $T = T_0$ for $\alpha = 1$, depending on $\delta$ and $w$. Green / yellow / red /  brown correspond to bounce with maximal $T$, bounce with minimal $T$, recollapse with minimal $T$ and recollpase with maximal $T$ respectively. Numbers from $1$ to $6$ represent different cosmological fates discussed in details in Figs. \ref{fig:case1}-\ref{fig:HTdeSitter} and in the text.} 
\label{fig:alpha1}
\end{figure}

The different numbers in the Fig. \ref{fig:alpha1} correspond to different scenarios of the evolution of the Universe, namely: 

\begin{enumerate}
    \item Region {\bf 1} defined as $w<0<\delta$ or $w>0$ and $0<\delta<-3+1/w$, corresponds to a cyclic Universe, where the maximal temperature is at the bounce and the minimal one at the recollapse.  
    \item In the region {\bf 2}, defined as $\delta>\max\{0,-3+1/w\}$ and $w>0$ or by $-6<\delta<-3$, one finds a recollapse followed by a rapid growth of the temperature. This leads to $x\to\infty$ at some finite time. We elaborate on this issue in the appendix.
    \item Region {\bf 3} defined as $0>\delta>3(1-3w)/(1+3w)$ and $w>-1/3$ is a recollapse within the range of $\delta$, for which $x_c$ (and a de Sitter evolution) could be in principle reached. Nevertheless, after the recollapse the temperature reaches $x=x_p=T_p/T_0$ (where $1<x_p<x_c$), for which $a_x(x_p)=0$. At $x=x_p$ the evolution of $x$ becomes discontinuous and therefore unphysical. For $x_p<x<x_c$ one finds growing Universe with de Sitter expansion in future infinity, similar to the dS bounce scenario. However, going back in time we will reach again $x_p$ with a discontinuity in the temperature so this is unphysical as well. 
    \item Region {\bf 4}, defined by $-3<\delta<\min\{0,3(1-3w)/(1+3w)\}$ is a single bounce. For $\delta>-2$, it is a  "normal bounce", at $x=1$ for which $x\to0$ for $t\to\infty$. So we have an initially contracting Universe followed by a bounce followed by a standard (non-accelerating) expansion with gradual decrease in temperature. Alternatively, for a different boundary condition $\rho(x_2)=0$ one finds the "dS bounce", with $x_2$ being the bounce temperature. 
    The temperature will asymptote to the critical temperature $x\to x_c$ as $t\to\infty$. One cannot obtain a cyclic Universe in this scenario, since $x\to 0$ (and therefore $\rho\to 0$) happens in future and past infinity. For $-3<\delta<-2$ the roles of $x=1$ and $x_2$ are simply switched:
    For $x=1$ one finds a dS bounce and for $x=x_2<1$ one obtains a ``normal'' bounce.
    Region {\bf 4} is the only part of the parameter space where one can obtain a "normal" bounce and reach the $x\to 0$ limit for $t\to\pm \infty$ and $a\to\infty$.  
    \item Region {\bf 5}, defined as $-3+\frac{1}{w}<\delta<-6$ is a recollapsing Universe with decreasing temperature. Since $\rho_u$ contains $T^{4+\delta}$ term and $4+\delta<0$, one finds $\rho \to \infty$ in a finite time. 
    \item Region {\bf 6}, defined as $w<0$ and  $\delta<\min\{-6,-3+\frac{1}{w}\}$ is a cyclic scenario, for which $x$ is minimal at the bounce and maximal at the recollapse. {We will denote such a scenario as an ``exotic'' cyclic Universe.}
\end{enumerate}

\subsection{Conditions for a Cyclic Universe and a single bounce} \label{sec:cycliccond}
The question of cyclic vs. a single bounce or recollapse lies in the existence of additional roots of $\rho(x\neq 1)=0$:
\begin{equation}
\alpha  x^{3 w-1} \left(\frac{4 (\delta +3)-3 (\alpha +1) (\delta +4) x^{\delta }}{\delta -3 \alpha  (\delta +4)}\right)^{w+1}-(\alpha +1) x^{\delta }+1=0 \, . \label{eq:conX}
\end{equation}
If $x=1$ is  the only solution of \eqref{eq:conX}, then the equations of motion only allow a single bounce or recollapse depending on the pressure $p(x=1)$. In such a case, the recollapse always leads to unstable Universe with $\rho \to \infty$.

In the case of multiple solutions of \eqref{eq:conX}, we first assume that $x=1$ is within our domain.  
In such a case we shall distinguish between two possible scenarios. In the first one the energy density is positive, between $x=1$ and $x=x_2$ so one can reach $x = x_2$ evolving from $ x = 1$. This case is simply a cyclic universe for which temperature is limited from below and above by two solutions of (\ref{eq:conX}). 
In the other case the energy density is negative between $ x = x_2$ and $ x = 1$ so one obtains two separate cosmological scenarios with the same values of parameters $\delta$, $w$ and $\alpha$, but different boundary conditions, namely $x = 1$ and $ x = x_2 $. These two scenarios have different allowed range of temperature and will evolve independently to a dS bounce or to a normal bounce with $x\to0$ for $t\to\pm\infty$.  
Examples of multiple solutions of \eqref{eq:conX} are shown in Fig. \ref{fig:rhox2}. 

Conditions for the existence of the cyclic Universe can be expressed in the following way: $\rho$ crosses $0$ at $x=1$. Therefore, it is positive for some range of $x$. One can be sure that a cyclic Universe exists, if $\rho<0$ for $x \to 0$ and $x\to\infty$. Considering equation \eqref{eq:rhoT}, the condition for a cyclic Universe if $\delta>0$ is significantly simplified and become independent of $\alpha$:
\begin{equation}
\boxed{
\begin{array}{rcl}
&\textit{cyclic:}& \\
&w<0<\delta& \\
\textit{OR} &\quad w>0 \quad \textit{and} \quad 0<\delta<-3+\frac{1}{w}&
\end{array}
}
\end{equation}
For $\delta<0$  there is some dependence on $\alpha$ as well:
\begin{equation}
\boxed{
\begin{array}{rcl}
&\it{cyclic:}& \\
&\alpha>\frac{1}{3}& \\
&w<0& \\
&\delta<\min\left\{\frac{-12\alpha}{(3\alpha-1)},-3+\frac{1}{w}\right\}&.
\end{array}
}
\end{equation}
Note that within those ranges one can obtain both ``normal'' and ``exotic'' cyclic scenarios, depending on the value of $\alpha$. For instance for $\alpha\ll\delta$ part of region {\bf 1} would be an ``exotic'' cyclic Universe labeled in brown (see Fig. \ref{fig:Phase-diagram}) for details. The condition for a single bounce is that $\rho\downarrow 0$ as $x\rightarrow 0$ from above. Considering \eqref{eq:rhoT} we get the following necessary and sufficient conditions:
\begin{equation}
\boxed{
\begin{array}{rcl}
&\textit{single bounce}& \\
\textit{necessary:} &-3\leq \delta\leq 0& \\
\textit{sufficient:} &-3\leq \delta \leq -2.
\end{array}
}
\end{equation}

\begin{figure}[]
\centering
\includegraphics[height=5cm]{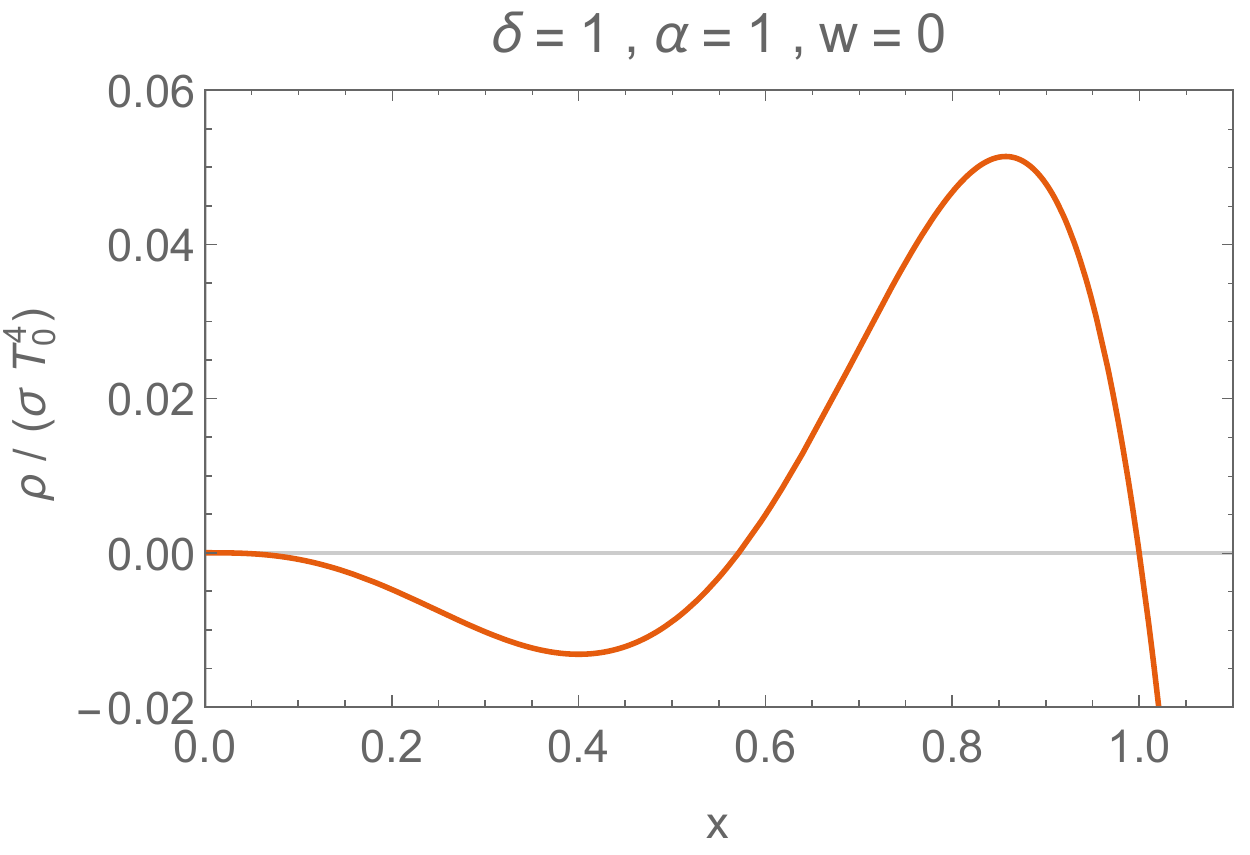}
\hspace{0.5cm}
\includegraphics[height=5cm]{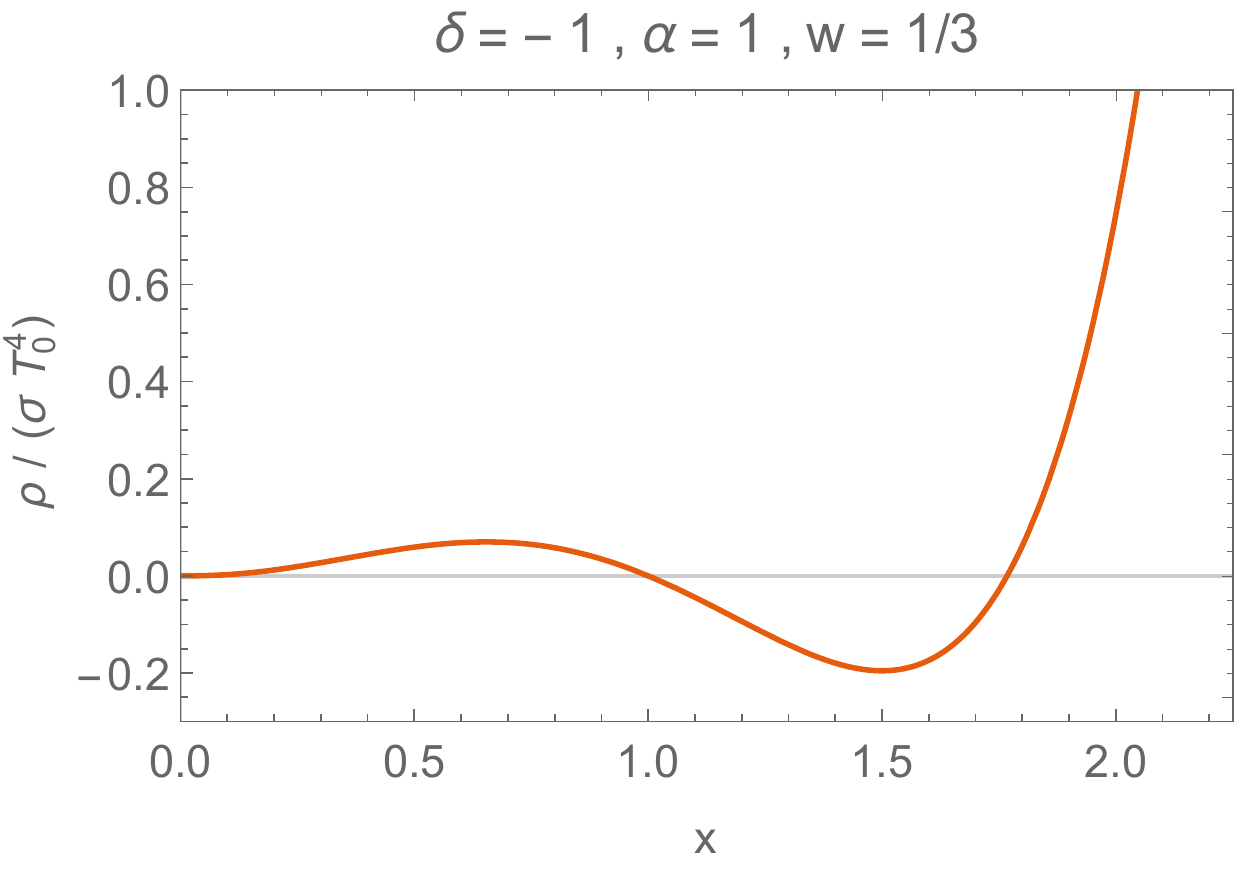}
\caption{\it Normalized $\rho(x)$ from  Eq. \eqref{eq:rhoT}.  Left panel shows the case of the cyclic Universe, from region {\bf 1}, for which $x=1$ at the bounce and $x\simeq 0.57$ at the recollapse. There are no other roots of $\rho=0$ continuously evolving to $\rho>0$.
Right panel shows two separated possible Universes with fixed values of the parameters $\alpha$, $\delta$ and $w$, but with different initial conditions from region {\bf 4}. For $0\leq x \leq 1$ one finds a single bounce at $x=1$ with $x\to 0$ for $t\to\pm\infty$. On the other hand, for $1.77\leq x <9/4$ one finds a dS bounce with the bounce at $x=1.77$ and the dS asymptotic temperature at $x_c=9/4$. This example shows how two consecutive solutions of Eq. \eqref{eq:conX} do not need to create a cyclic scenario. Both scenarios from the right panel are presented in Fig. \ref{fig:HTdeSitter}. }
\label{fig:rhox2}
\end{figure}
As explained before, each single bounce solution will contain both a dS branch and a "normal" branch. For the rest of the section we discuss the different viable scenarios: Cyclic universes, dS Bounce,  and "normal bounce" - bounce followed by deceleration. Finally we shall discuss the special case where the bounce does not occur at an extremum of the temperature. The recollapsing scenarios that result in future singularities have an interesting singularity structure. We relegate the discussion of the various instabilities and singularities to the appendix. We would like to stress that the non-singular solutions were obtained from dynamically solving the equations of motion with given initial conditions, and are not a result of additional external constraints.

\subsection{Cyclic solutions}\label{sec:cyclic}
The condition for a cyclic scenario is that apart from $x=1$, there exists another root of the Eq. \eqref{eq:conX} and that a positive energy density $\rho\geq 0$ interpolates between them.  

In addition, for a realistic cyclic scenario one requires a very large range of temperatures, i.e. $x_2 \ll 1$ or $x_2\gg1$ (for a bounce or recollapse at $x=1$ respectively). This can be easily obtained. For example, for small $\delta \ll 1$ the temperature at the bounce is $x=1$  (region {\bf 1}\,,\,green) and at recollapse:\footnote{Note that the result is $w$-independent. Nevertheless, one must assume $w < 1/3$ in order to obtain a bounce at $x=1$ and recollapse at $x=x_2$.} 
\begin{equation}
x_2 \sim \left(1+\alpha\right)^{-1/\delta} \, . \label{eq:x2app}
\end{equation}
{$x_2$ is the lower bound of $x$. In this case one can obtain simple solutions of Friedmann equations around the recollapse}
\begin{eqnarray}
x &\simeq& x_2 \left( 1+\frac{\delta \, \sigma}{12}T_0^4x_2^4 t^2 \right) \, . \\
H &\simeq& - \delta \, \sigma T_0^4 x_2^4 t \, ,
\end{eqnarray}
{where $t=0$ denotes the moment of the recollapse.} 

A big ratio between minimal and maximal temperature of the Universe may also be obtained in the red part of region {\bf 1}. In such a case, for  $\delta \lesssim -3+1/w$ one finds
\begin{equation}
x_2 \sim  \left(\frac{1}{\alpha }+1\right)^{\frac{1}{(\delta +3) w-1}} \left(-3 \frac{(\alpha +1) (\delta +4)}{\delta -3 \alpha  (\delta
   +4)}\right)^{-\frac{w+1}{(\delta +3) w-1}} \, .\label{eq:x2app2}
\end{equation}
Approximations presented above are very consistent with numerical values of $x_2$, as shown in Fig. \ref{fig:x2}.

\begin{figure}[H]
\centering
\includegraphics[height=5cm]{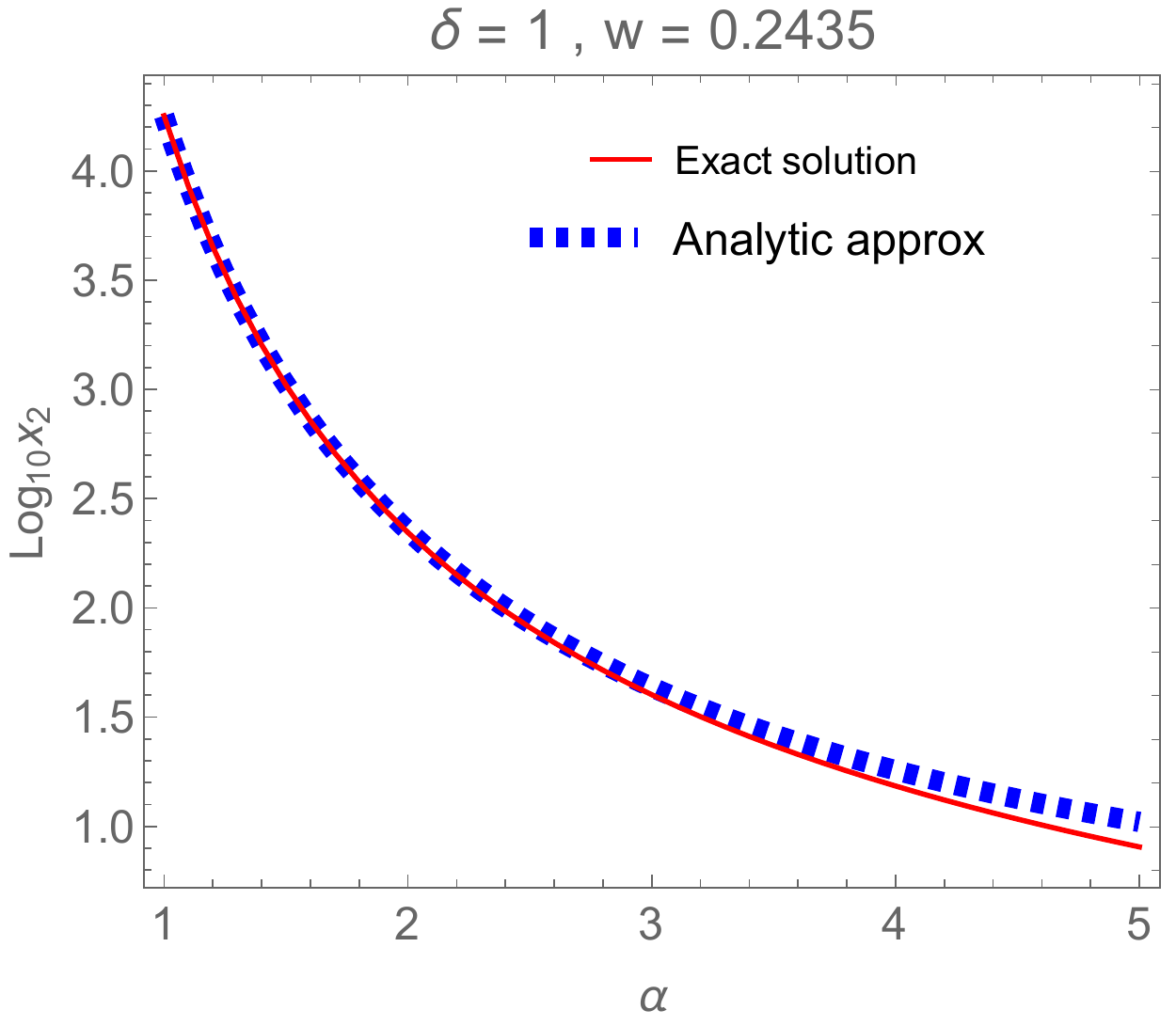}
\hspace{0.5cm}
\includegraphics[height=5cm]{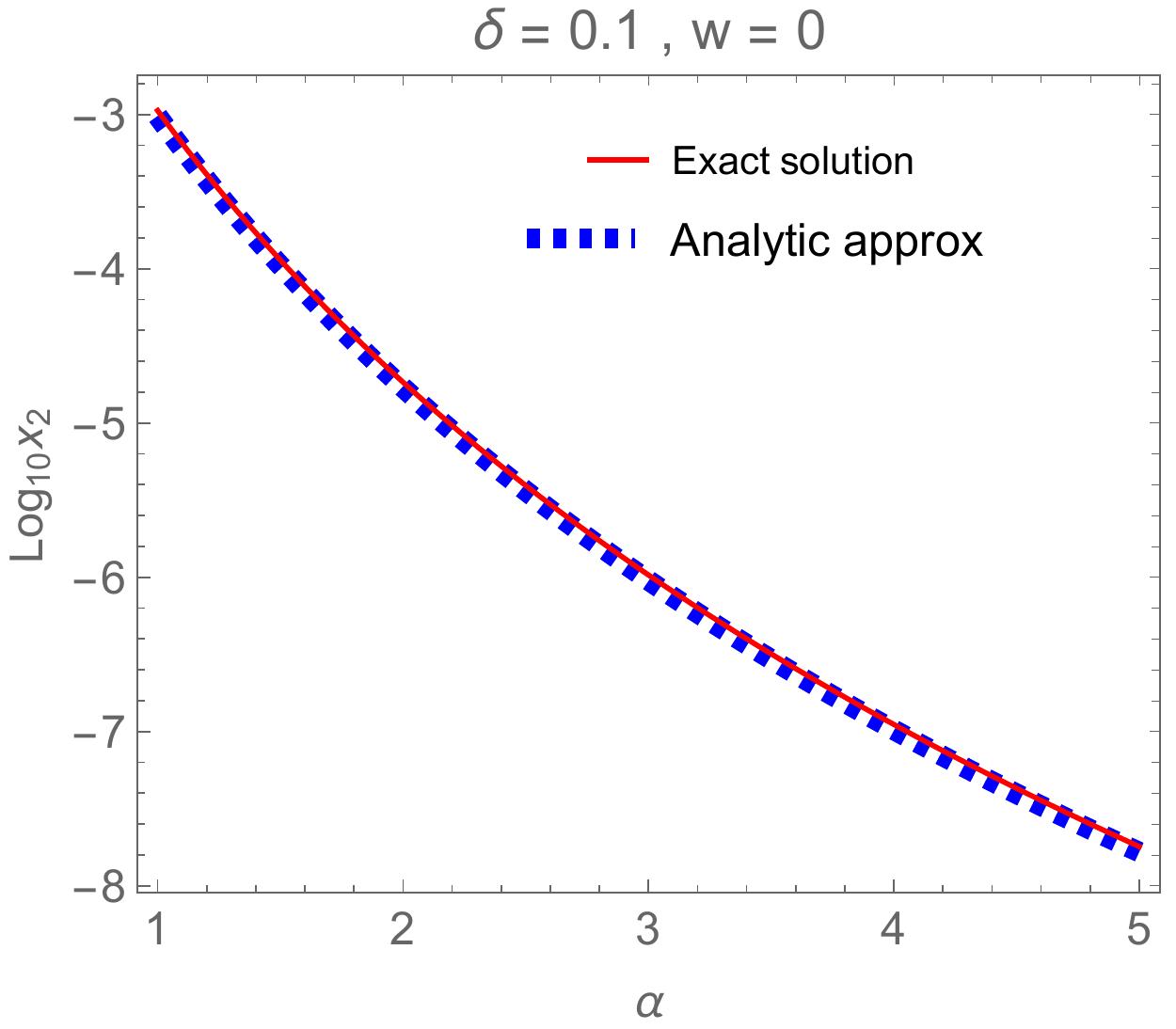}
\caption{\it Left and right panels show $x_2$ in region {\bf 1}. Solid and dotted lines represent exact solutions and analytical approximations of the Eq. \eqref{eq:conX}. Note that approximate results from Eqs. \eqref{eq:x2app} and \eqref{eq:x2app2} fit the exact values of $x_2$ very well.} 
\label{fig:x2}
\end{figure}

An example of the cyclic scenario of region {\bf 1} is given in Figure \ref{fig:case1}. We present the evolution of the Hubble parameter $H$, the temperature $T$ in the left panel and the NEC evaluation on the right panel. NEC is violated only around the bounces. The maximal temperature is occurring at the bounces, and the minimal temperature at the recollapses.\footnote{ In the plotted example $T_{max}/T_{min}=10^4$, however the only limitation is the numerical integration and there is no obstacle to having this ratio arbitrarily large by considering smaller and smaller $\delta$  for negative $w$, or $\delta\lesssim -3+1/w$ for $w>0$. For $\alpha=1$, a plausible example in the Banks-Zaks case is $\delta\sim 0.03$ that will result in $T_{max}/T_{min}\simeq 10^{10}$ and $\delta =0.01$ to $T_{max}/T_{min}\sim 10^{31}$. {The duration of a cycle is proportional to $C^{1/\delta}$, where $C>0$ is some $\alpha-$dependent constant. For $\delta\leq 0.01$ and $\alpha=1$ the duration of a cycle is longer than the age of the Universe.}}
\begin{figure}[H]
\centering
\includegraphics[height=6cm]{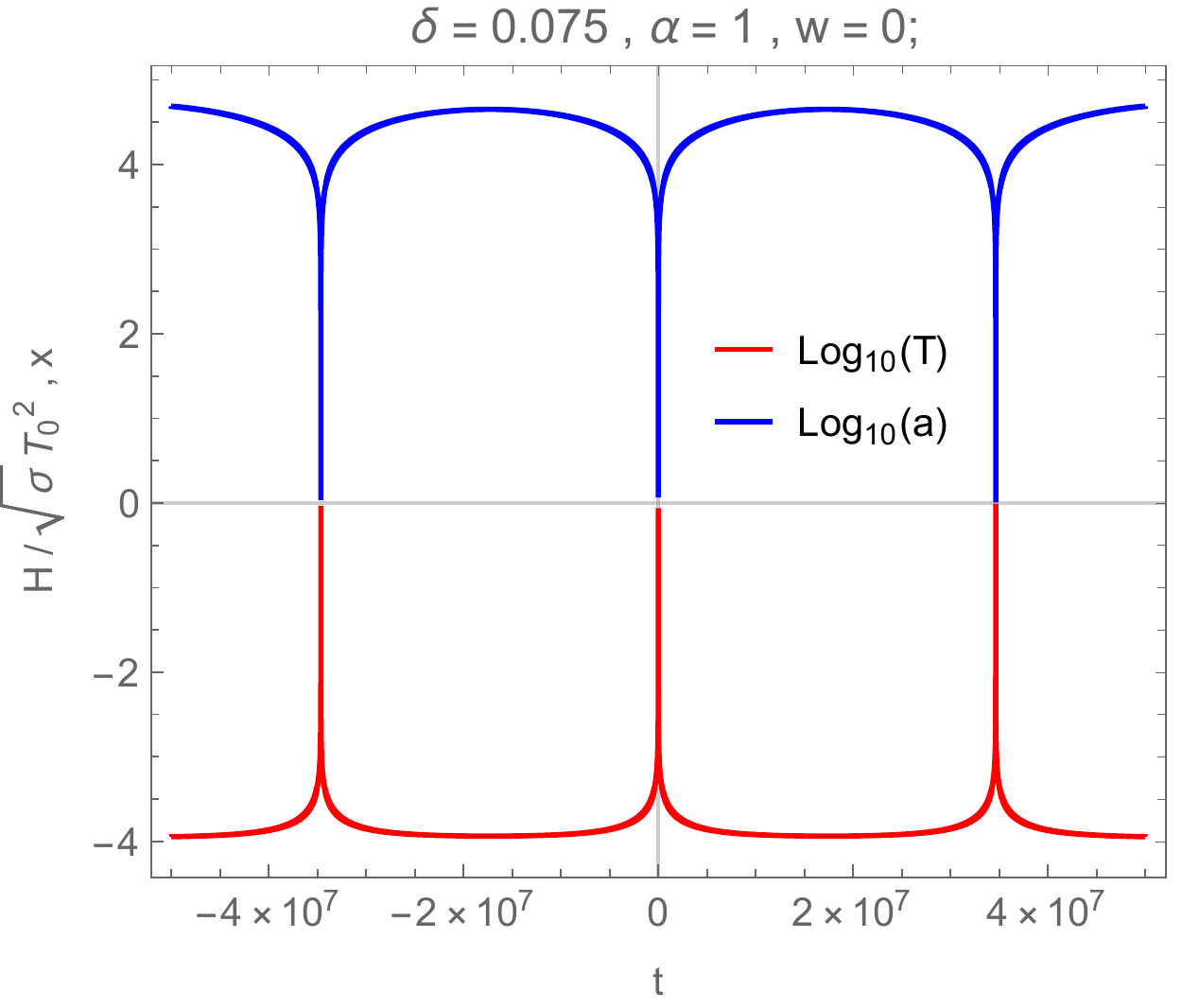}
\hspace{0.5cm}
\includegraphics[height=6cm]{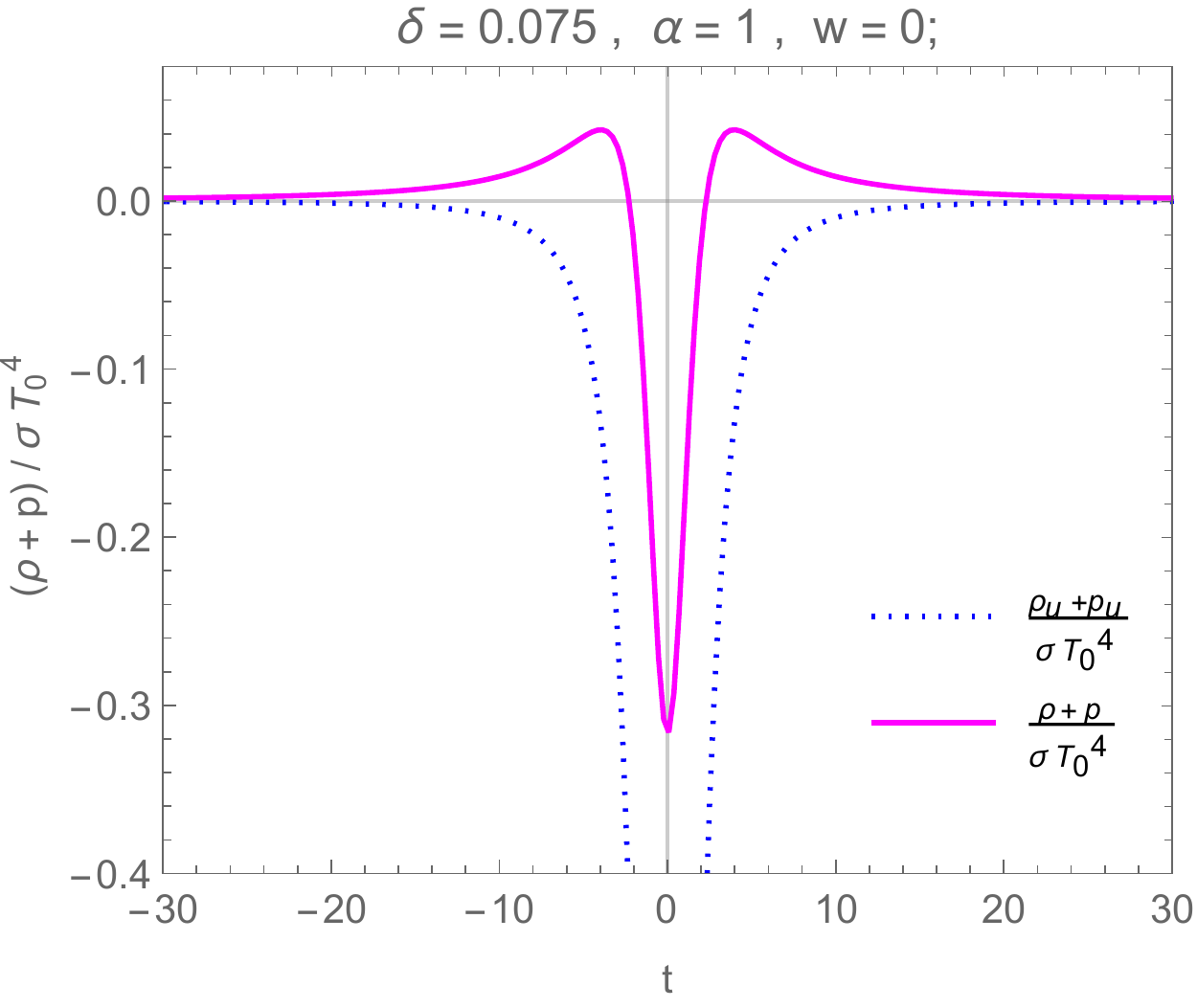}
\caption{\it Left panel: Evolution of the normalized temperature $x=T/T_0$ (red) and the normalized scale factor $ a(t)$ (blue). This is an example of region {\bf 1} from  Fig. \ref{fig:alpha1}. {Time is expressed in Planck units.} One obtains a cyclic Universe, for which $T = T_0$ i.e. $x=1$ is a maximal temperature and this temperature is obtained at the bounce. Right panel:  The normalized value of the NEC. Solid magenta line corresponds to the total energy density and pressure, while the dashed blue line corresponds to the unparticles only. NEC is violated only around the bounces.}
\label{fig:case1}
\end{figure}
 
The second case of the cyclic Universe occurs in the brown and the yellow colors of region {\bf 6} of Fig. \ref{fig:alpha1}, and is plotted in Fig. \ref{fig:case6}. It involves a cyclic solution but with minimal temperature at the bounce and maximal at the recollapse. The only difference between different colors in this case is the fact that for brown and yellow of region {\bf 6} one obtains a recollapse and bounce at $x=1$ respectively.
\begin{figure}[H]
\centering
\includegraphics[height=6cm]{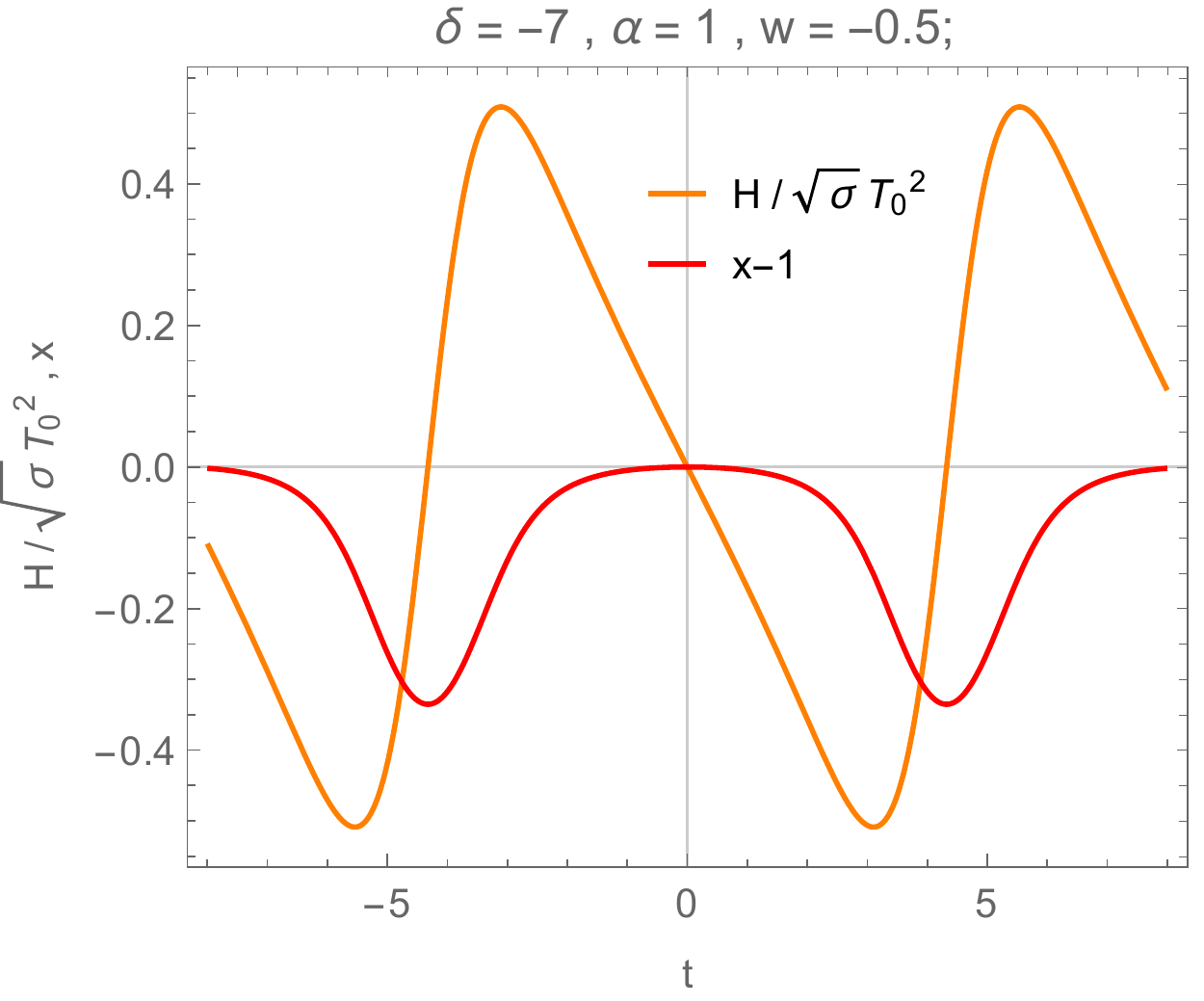}
\includegraphics[height=6cm]{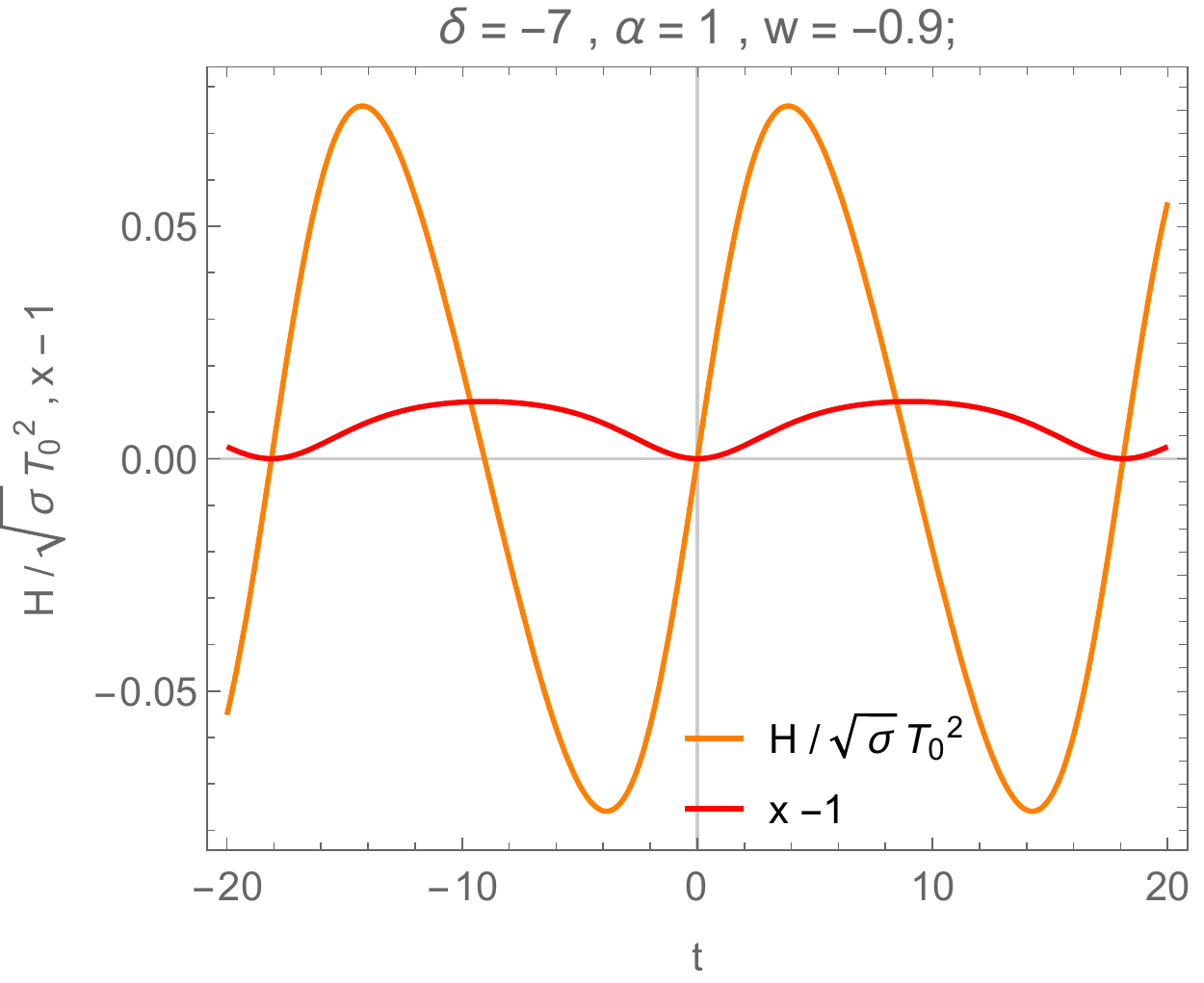}
\caption{\it Left and right panels show examples of the evolution of the normalized Hubble parameter and normalized temperature for region {\bf 6} of Fig. \ref{fig:alpha1}. {Time is expressed in Planck units.} The Universe obtains the minimal temperature at the bounce and the maximal temperature at the recollapse. To improve the presentation in the left panel $x-1$ is plotted. The actual temperature is obviously always positive.} 
\label{fig:case6}
\end{figure}

\subsection{Single bounce solutions}\label{sec:single}
Multiple solutions of Eq. \eqref{eq:conX} do not need to lead to the cyclic scenario. This will occur from dynamically solving the equations of motion if the different roots of $\rho=0$ are not connected by $\rho>0$ region. Depending on the initial conditions, we will either have a single dS bounce as in the pure unparticle case, or we shall have a "normal bounce". This corresponds to regions {\bf 4} yellow and green respectively. An example of both possibilities is depicted in Fig. \ref{fig:HTdeSitter}. The example of Fig. \ref{fig:HTdeSitter} is the manifestation of the two branches demonstrated in right panel of figure \ref{fig:rhox2}. Let us note that  contrary to other scenarios discussed here, the "normal bounce" does not have a lower bound on the temperature {other than $T>0$}. 

\begin{figure}[H]
\centering
\includegraphics[height=5.0cm]{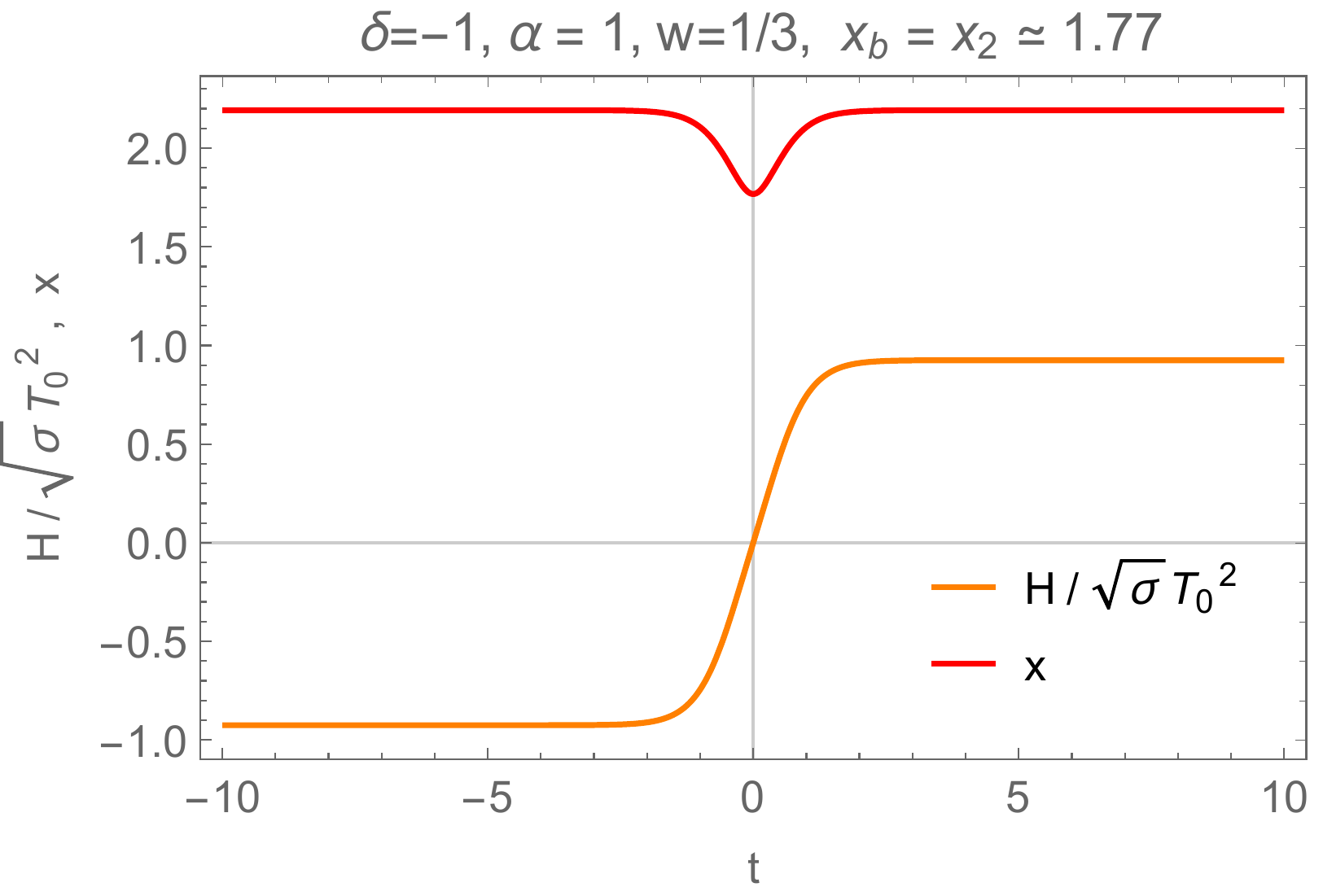}
\hspace{0.3cm}
\includegraphics[height=5.0cm]{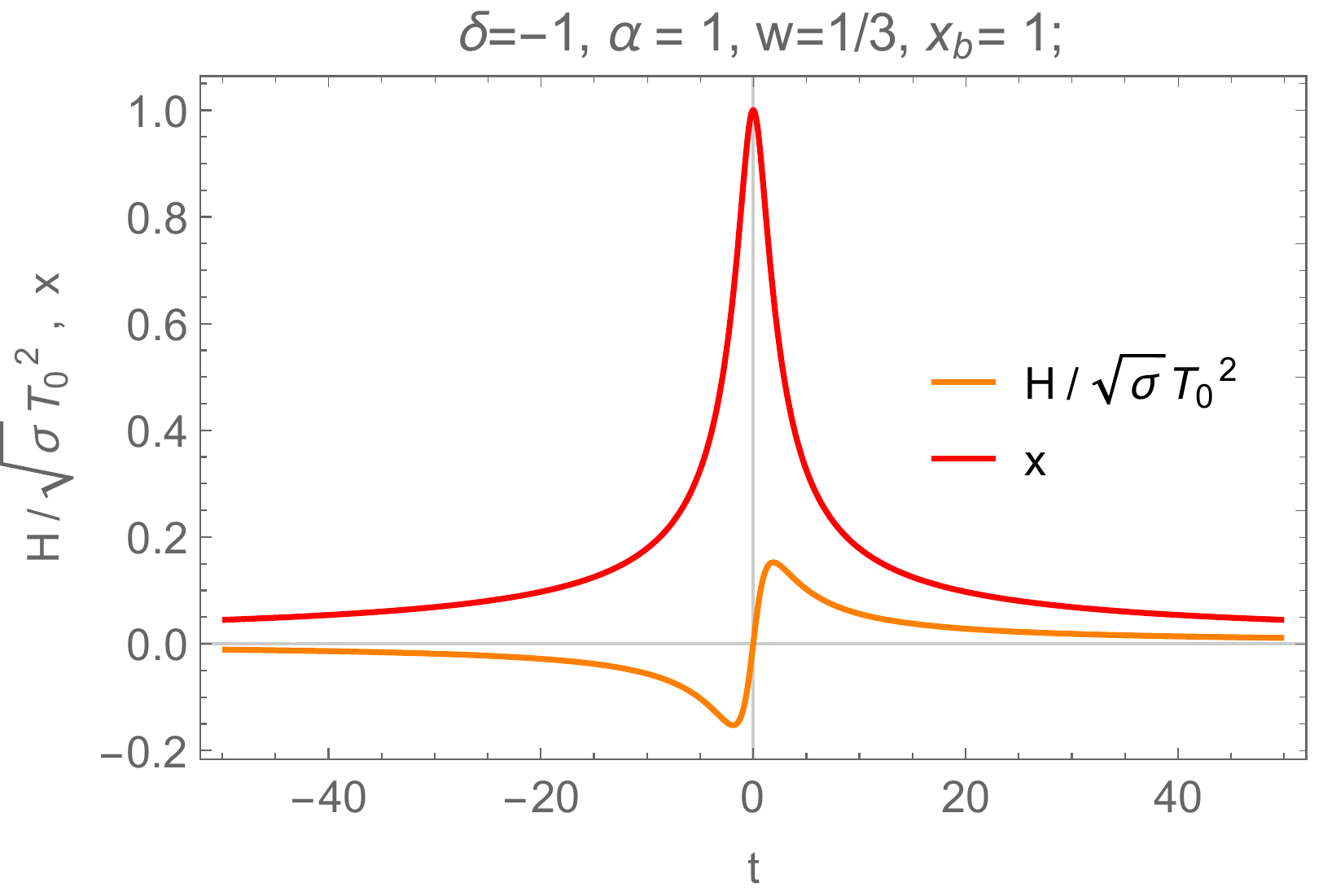}
\caption{\it Normalized Hubble parameter and normalized temperature as a function of time. Numerical solutions for $\alpha = - \delta = 1$ and $w=1/3$. {Time is expressed in Planck units.} Left and right panels represent bouncing scenarios for a bounce at  $x = x_2  \simeq 1.77$ and $x = 1$ respectively. The left panel shows a dS Bounce, while the right panel demonstrates a "normal bounce" followed by a decelerated evolution. Both correspond to region {\bf 4} of Fig. \ref{fig:alpha1}. Note how different values of $x$ at the bounce lead to radically different scenarios for the Universe.
}
\label{fig:HTdeSitter}
\end{figure}

\subsection{The special case of $\delta = -\frac{4\alpha}{1+\alpha}$} \label{sec:special}

As noticed, for $\delta = -\frac{4\alpha}{1+\alpha}$ one can obtain $H(x=1) = 0$ without $\dot{x}(x=1) = 0$. For this particular value of $\delta$ one obtains $\dot{H}_0 \simeq \frac{3-\alpha}{\alpha +1}\dot{x}_0^2${, which gives a bounce for any $\alpha<3$}.
    For $\alpha = 1$ one finds $\delta = -2$, which in Fig. \ref{fig:alpha1} is a borderline between yellow and green regions. Indeed, the case of $\delta = -\frac{4\alpha}{1+\alpha}$ is a transition between those two regions of the parameter space, since it generates a transition between de Sitter Universe and decelerated evolution of a scale factor and breaks the symmetry around each bounce. As a result, one can have a slow contraction phase followed by a dS phase. This is in contrary to the previous analysis where a future dS phase with $H_c$, implied a previous exponential contraction with similar negative $-H_c$ prior to the bounce. Having $\dot{x} \neq 0$ at the bounce is possible in this case because for $\alpha > 1/3$ and any $w$ or for $\alpha<1/3$ and $-1<w<(7+3\alpha)/(9-3\alpha)$ the energy density obtains a minimum at $x=1$. Thus, one obtains $\rho \geq 0$ for all $x$. An example of the evolution of $H$ and $T$ in this special case is presented in Fig. \ref{fig:special}.
\begin{figure}[H]
\centering
\includegraphics[height=5.5cm]{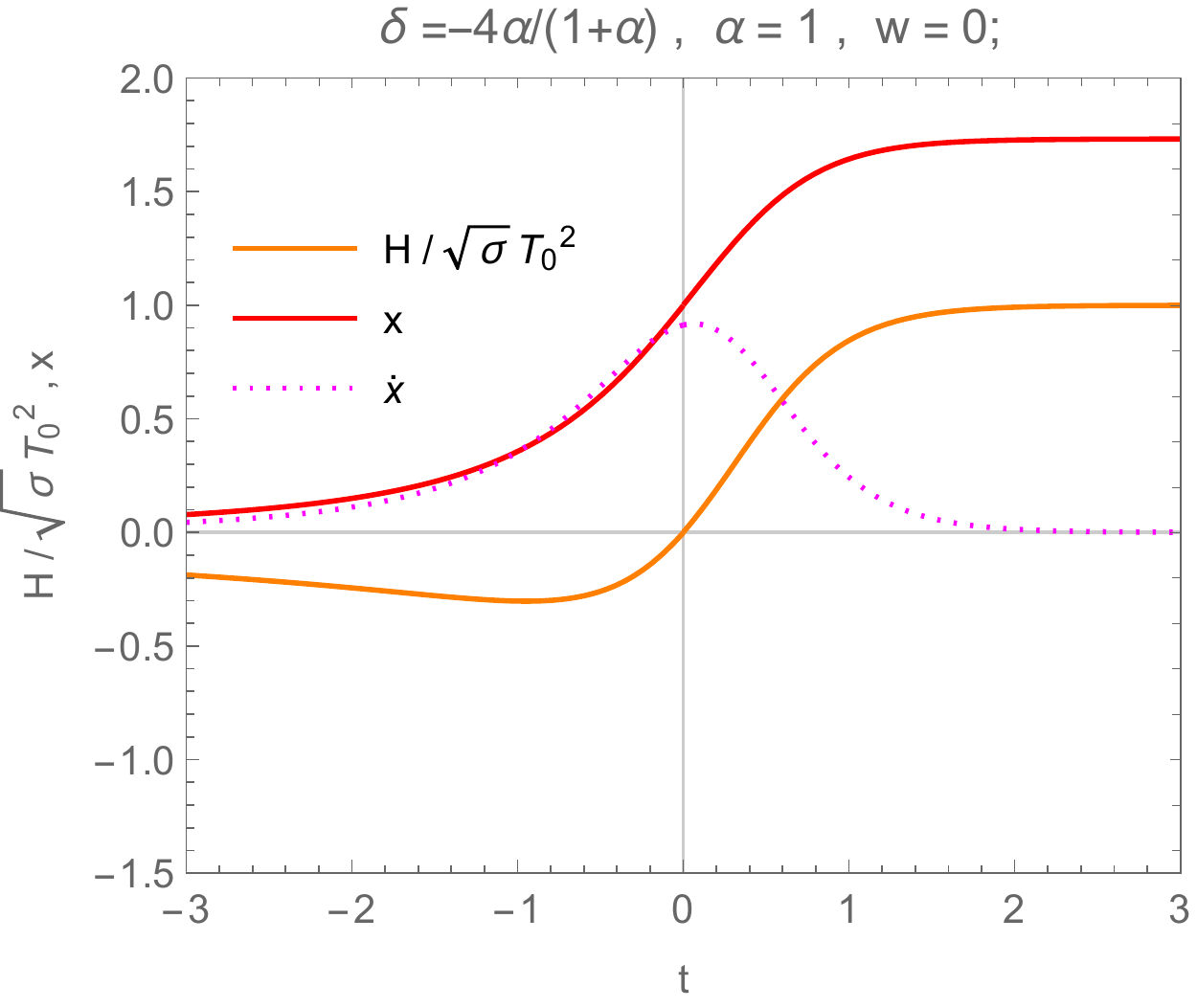}
\hspace{0.5cm}
\includegraphics[height=5.5cm]{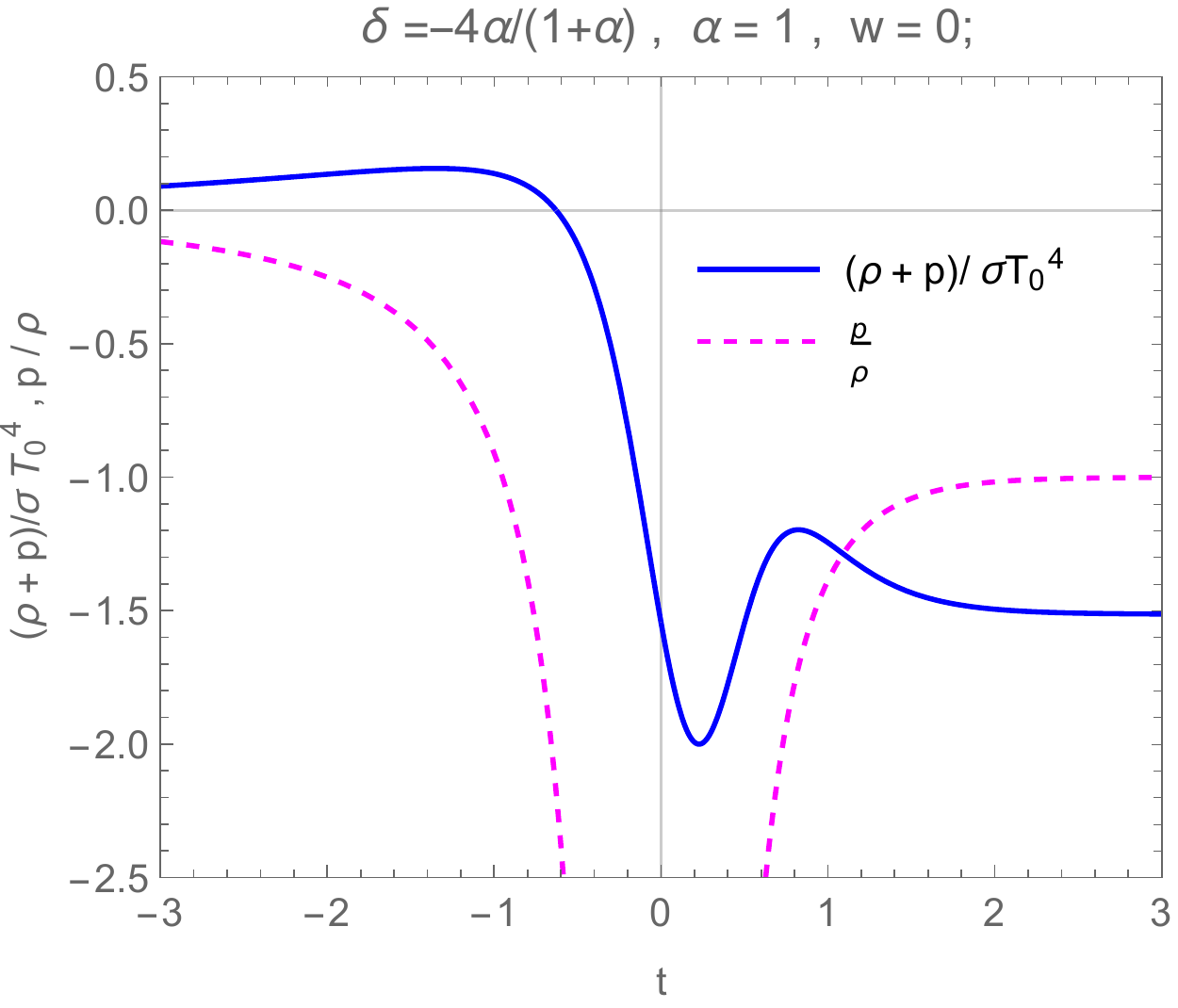}
\hspace{0.3cm}

\caption{\it Left panel: The normalized Hubble parameter, normalized temperature $x$ and its time derivative $\dot{x}$ as a function of time for $w=0,\, \alpha =  1\,,\,  \delta =  -\frac{4\alpha}{1+\alpha}=-2$ \label{fig:special}, which is a border-line between the yellow and green parts of region {\bf 4}. In this case $\dot{x}(x=1) \neq 0$, so the temperature does not obtain an extremum at the bounce. In this unique case one finds a transition between a decelerated contraction and a de Sitter-like exponential growth of the Universe before and after the bounce respectively. By choosing different initial conditions one could obtain a reverse scenario, in which exponential contraction would be followed by the decelerated growth of the Universe. Right panel: Evolution of normalized $\rho + p$ and $\frac{p}{\rho}$ (solid and dashed lines respectively). Note that NEC is violated around the bounce and throughout the dS phase. {In both panels time is expressed in Planck units.}}

\end{figure}
To conclude, the Universe filled with unparticles and a perfect fluid offers a rich spectrum of cosmological scenarios that are never singular. The allowed solutions include cyclic models as well as "symmetric" dS Bounce or a "normal bounce" depending on the parameters of the model. For the specific case where the temperature is not extremal at the bounce, one also obtains asymmetric bounces. In each scenario, there is a domain of the parameters where the range of temperatures is parameterically large and viable.  Perhaps the most interesting case is the small $0<\delta \ll 1$ limit. This small anomalous dimension is sufficient to discard the singularity and produce a viable scenario, while being very close to the conformal point. It therefore deserves a closer inspection, that we turn to next.  

\section{Revision of the Banks-Zaks thermal average} \label{sec:fixedmu} 

In previous sections we have assumed that the renormalization scale $\mu$ is equal to $T$. In this section we are taking a more conservative approach by taking a fixed renormalization scale $\mu$ and varying temperature $T$. From dimensional analysis  \eqref{ev} is changed to:
\begin{equation}
\langle N\left[ F_a^{\mu} F_{a \, \mu \nu}\right]\rangle = C\frac{ T^{4 + \gamma }}{\mu^{\gamma}} 
\end{equation}
where $C$ is dimensionless and $\gamma$ is the anomalous dimension of the operator \cite{Weinberg:1976xy}. Since we are interested in physics at large distances, i.e. low energies, we are considering $T \ll \mu$. As a result the trace of the energy momentum tensor near the conformal point $g_{*}\gg u\mu^a$ will be
\begin{equation}
\theta^{\mu}_{\mu}\simeq \frac{C a u \mu^{a-\gamma}}{2g_{*}}T^{4+\gamma}+\mathcal{O}\left(\frac{u\mu^a}{g_{*}}\right)\equiv \tilde{A}T^{4+\gamma}
\end{equation}
Notice that the power of temperature is different. Calculation of the anomalous dimension of the operator gives:
\be
\gamma=\Delta-d=a+d_S-d=a
\ee
where $\Delta$ is the scaling dimension of the operator, $d_S$ is the engineering dimension of the operator and $d$ is the engineering of the operator. Since $d_S=d=4$, we get $\gamma=a$.
As a result, the functional form of the $\theta^{\mu}_{\mu}$ of the pressure and energy density of the Banks-Zaks/unparticles fluid is unchanged. However, now $\delta\equiv a$, so $\delta$ is now limited to $1 \gg \delta\geq0 $. Since at the level of equations the $\mu \neq T$ case is effectively just a subset of $\mu = T$, one can read off possible cosmological scenarios for $\mu \neq T$ from the analysis presented in sections \ref{sec:uonly},\ref{sec:fluid}.

As shown in Sec. \ref{sec:uonly}, in the case of unparticles only, such a range for $\delta$ cannot lead to a bounce. It either describes a hot Big Bang scenario, or it has a discontinuity in $H$ and $T$, see again Figure \ref{fig:alpha=0}. However, in the case of unparticles with a  perfect fluid one can still obtain bouncing and recollapsing solutions. From the discussion in sec. \ref{sec:cycliccond} and from Fig. \ref{fig:alpha1} we realize that the only viable non-singular solutions will correspond to cyclic ones, while single bounces are precluded. Having a small parameter simplifies the analysis considerably. The condition for a cyclic scenario becomes $w<1/3$. For $\alpha>\mathcal{O}(\delta)$ the bounce will occur at maximal temperature and the recollapse at the minimal one (Region {\bf 1}). However, if there is further hierarchy, $\alpha \ll \delta \ll 1$, the "exotic cyclic" scenario with bounces at minimal temperature may occur. 

Examples of the phase diagram depending on the values of $\alpha $ is depicted in Figure \ref{fig:Phase-diagram}. The plot shows all parametric dependence of theory. Once again, the striking feature is that the small anomalous dimension is sufficient to produce a non-singular universe with phenomenologically viable range of temperatures, and without using the scalar field paradigm. Of particular interest are again the regions {\bf 1} with $w<1/3, \, 0<\delta \ll 1$ that give a realistic cyclic scenario with a huge range of tempertures during the evolution.
\begin{figure}[H]
\centering
\includegraphics[height=6.1cm]{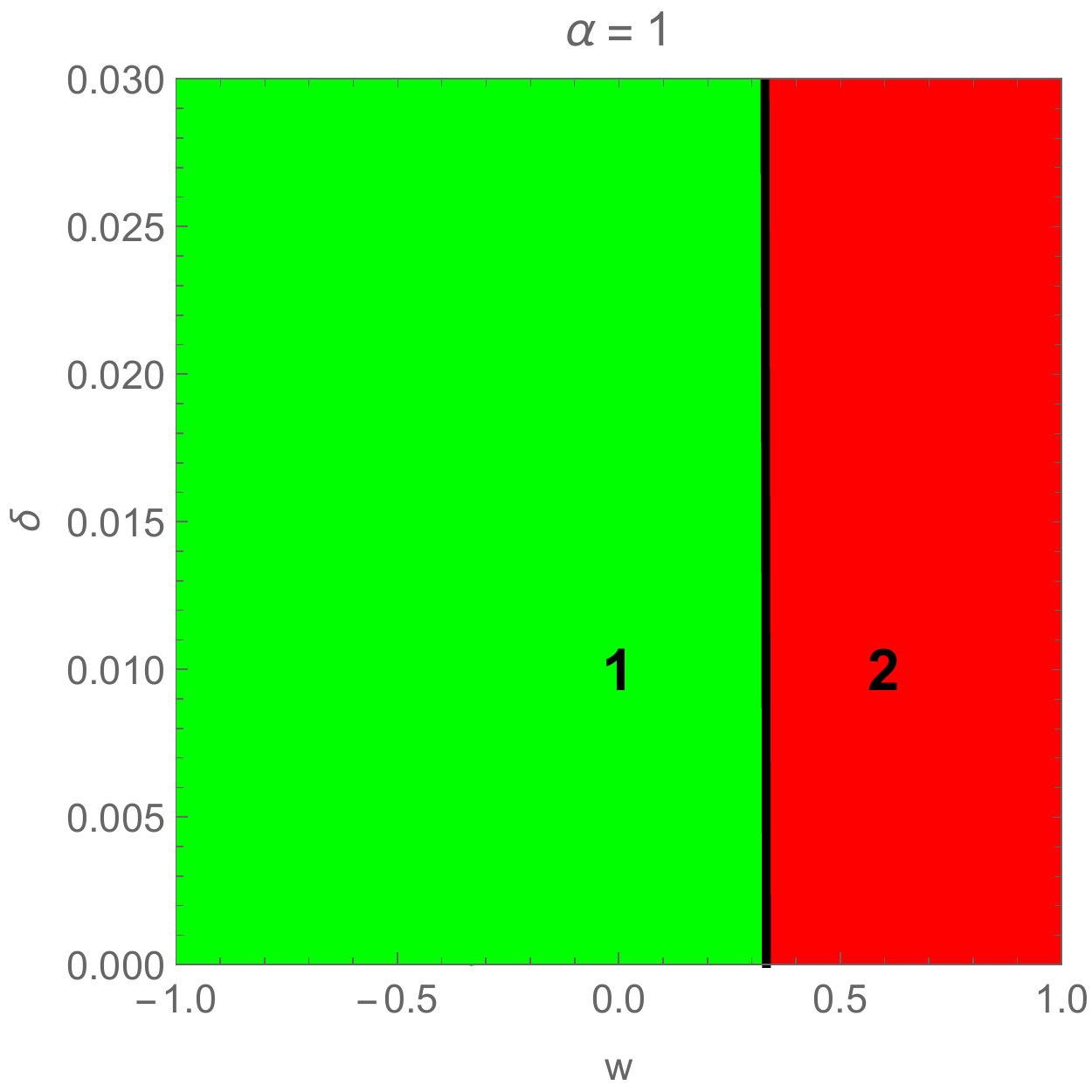}
\hspace{0.5cm}
\includegraphics[height=6.1cm]{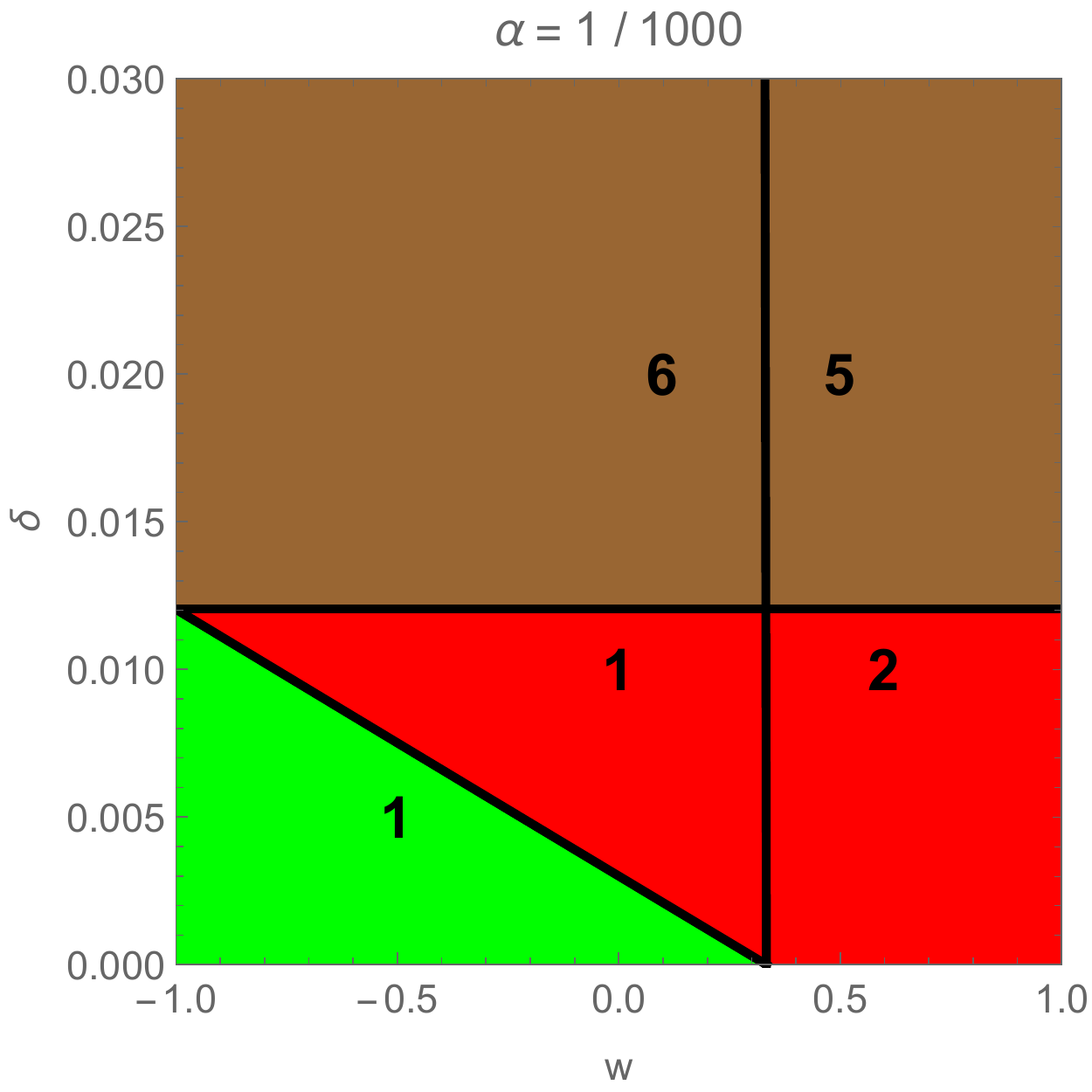}
\caption{\it Phase diagram for $\alpha = 1 $(left panel) and $\alpha = \frac{1}{1000}$ (Right panel). Green, red, brown represent the bounce with the maximum of the temperature, recollapse with the minimum of the temperature and recollapse with the maximum of the temperature respectively. The numbers of the regions represent the same physical cases of the evolution of the Universe as in the Fig. \ref{fig:alpha1}. Hence, {\bf 1} represents a cyclic scenario with bounces at maximal temperatures, {\bf 6} exotic cyclic scenario with bounces at minimal temperatures, and {\bf 2,5} represent solutions that become singular at finite time. }
\label{fig:Phase-diagram}
\end{figure}

The existence of a small parameter also allows us to write analytic approximations to the different quantities, and solve the equations of motion analytically. The various expressions are given as follows and can be easily used for future analyses.

{Let us note that for $B=0$ or $\delta=0$ unparticles are fully equivalent to the standard radiation. Therefore, in the $|\delta| \ll 1$ and $ |B| \ll 1$ regime one can consider unparticles as a radiation-like fluid with small corrections from non-zero values of $B$ and $\delta$. In $|\delta| \ll 1$ approximation one finds 
\begin{eqnarray}
\rho_u &\simeq& \sigma\, T^{4} \left(1 + \frac{B}{\sigma}\left(1 + \delta \log T\right)\right) \label{eq:small delta rho} \, ,\\
p_u &\simeq& \frac{1}{3} \sigma\, T^{4}\left( 1 + \frac{B}{\sigma}\left( 1 +\frac{\delta}{3} - \delta  \log T\right)\right) \label{eq:small delta p}\, .
\end{eqnarray}
Using Eqs (\ref{eq:small delta rho}) and (\ref{eq:small delta p}) one finds following expressions for $T(t)$, $\rho_u(t)$ and $ a(t)$
\begin{eqnarray}
T(t) &\simeq & T_{i} \left(\frac{t_i}{t}\right)^{\frac{1}{2}}\left( 1 + \frac{B\, \delta}{8 \left(\sigma + B\right) } \log \frac{t}{t_i}\right)\, , \label{eq:smallT} \\
\rho_u(t) &\simeq & \rho_{u_{i}} \left(\frac{t_i}{t}\right)^2 \left( 1 + \frac{B\, \delta^{2}}{4 \sigma} \log \frac{t}{t_i}\right)\, , \label{eq:smallrho} \\
a(t) &  \simeq & a_{i} \, \left( \frac{t}{t_i}\right)^{\frac{1}{2}}\left( 1 + \frac{B\, \delta}{24 \left(B + \sigma\right)} \log \frac{t}{t_i}\right)\, , \label{eq:smalla}\\
H &\simeq& H_i \left(\frac{t_i}{t}\right) \left( 1 + \frac{B \delta^{2}}{8\sigma} \log \frac{t}{t_i}\right) \, . \label{eq:smallHB}
\end{eqnarray}
where $T_i$,  $\rho_{u_{i}}$ and $ a_i$ are some initial values of respective quantities at some initial time $t_i$. Indeed, the leading order in Eqs (\ref{eq:smallT}- \ref{eq:smalla}) behaves like radiation. 
Similarly, considering $|B| \ll 1$ also gives the unparticles as a radiation like fluid with small corrections from non zero B. In this approximation one finds
\begin{eqnarray}
T(t) & \simeq & T_i \, \sqrt{\frac{t_i}{t}}\left( 1 - \frac{B}{\sigma^{\frac{5}{4}}} \left(\frac{t_i}{t}\right)^{\frac{\delta}{4}}\right) \label{eq: smallBT}\, , \\
\rho_u(t) &\simeq& \rho_{u_{i}} \left(\frac{t_i}{t}\right)^{2}\left( 1 + \frac{B}{\sigma^{\frac{5}{4}}} \left(\frac{t_i}{t}\right)^{\frac{\delta}{2}}\right) \label{eq: smallBrho}\, ,\\
a(t) & \simeq & a_{i}\, \sqrt{\frac{t_i}{t}}\left( 1 - \frac{B}{12 \, \sigma^{\frac{5}{4}}} \left(\frac{t_i}{t}\right)^{\frac{\delta}{4}}\right) \label{eq: smallBa}\, , \\
H(t) & \simeq & H_{i}\, \left( \frac{t_i}{t}\right) \left( 1 + \frac{B\,}{2\, \sigma^{\frac{5}{4}}}\left(\frac{t_i}{t}\right)^{\frac{\delta}{2}}\right)\, . \label{eq:smallBH}
\end{eqnarray}
{Obviously the approximate solutions (\ref{eq:smallT}-\ref{eq:smallBH}) are valid only in limited time range. Whenever $\delta\log(t/t_i)$ or $B(t_i/t)^{\delta/4}$ becomes much bigger than unity one should include higher order corrections like $\delta^2\log^2(t/t_i)$. Furthermore, one can extend this analysis into unparticles plus radiation scenario, which requires modifying $\sigma$ to include additional relativistic degrees of freedom.}


\section{Discussion} \label{sec:summary} 
We have investigated the possibility of obtaining a non-singular Universe filled with unparticles or unparticles + perfect fluid. The unparticles only case results in the exponentially contracting universe followed by a bounce and then exponentially expanding universe for $B>0 $ and $-3<\delta <0$. We called this scenario a dS Bounce. The scenario naturally has an inflationary phase, but without a Big Bang singularity! Another interesting case is the asymptotic dS phase with empty Minkowski Universe as an initial condition, (pink region of Fig. \ref{fig:alpha=0}). For any other values of parameters one encounters either Big Bang singularity or instabilities. For the dS Bounce, the NEC is always violated and a bounce happens at the minimum of the temperature. The temperature is bounded from below by the temperature at the bounce ($T_b$) and from above  by  the temperature of the de Sitter expansion ($T_c$).  Besides the case of $ \delta + 3 \ll 1 $, $T_b$ and $ T_c$ are of the same order of magnitude. Exponential growth of the scale factor mimics the cosmic inflation with no graceful exit. Inflation generated by unparticles  resembles the so called constant roll inflation, with the difference being  that the slow roll parameter $\epsilon $ is negative.

\paragraph{} In section \ref{sec:fluid} we have generalized this analysis to the case of unparticles with a perfect fluid.  As a result, one finds a rich spectrum of scenarios like bounces and recollapses with minima or maxima of temperature. In addition, both bouncing and recollapsing Universes may lead to a cyclic scenario. We have defined general sufficient conditions for the existence of the cyclic Universe as $0<\delta<-3+1/w$ (where the upper bound is taken for $w>0$) or $\delta<\min\{-12\alpha/(3\alpha-1),-3+1/w\}$ for $w<0$ and $\alpha>1/3$. Similarly we have derived necessary and sufficient conditions for a single bounce $-3\leq \delta\leq 0$ and $-3\leq \delta \leq -2$ respectively. In this setting the bounce / recollapse appears as the extremum of the temperature. The allowed range of temperature may be quite broad comparing to the unparticles only scenario, which makes the model more realistic.  
In the single bounce case, one may obtain different temperatures of the bounce for the same values of parameters, which may lead to radically different cosmological scenarios. An example of such a case is shown in Fig (\ref{fig:HTdeSitter}), where we have presented a de Sitter-like scenario (similar to the one from the unparticles only case) together with a more conventional bouncing model with the decelerated expansion of the Universe after the bounce. 

\paragraph{} In section \ref{sec:fixedmu} the case of the fixed renormalization scale (i.e. $ \mu \neq T $) has been investigated. This approach seems more appropriate from the field theory point of view. For this case expressions for energy density and pressure of unparticles remain unchanged. Nevertheless, the parameter space is constrained $ 0<\delta \ll 1$. Bouncing scenarios cannot be  realized in this case for unparticles only. For unparticles + fluid all regions presented in Fig \ref{fig:alpha1} can be obtained except the yellow region. Hence, in the $0<\delta \ll 1$ case the temperature is always maximal at the bounce. Viable non-singular scenarios here are only cyclic universes, in most cases with maximal temperature at the bounce and minimal at the recollapse. 

To recap, according to our analysis the main promising scenarios are the "genesis" from Minkowski space in section \ref{nonbounce}, the cyclic scenarios of $\delta \ll 1$ or $\delta\lesssim -3+1/w$ of region {\bf 1} in section \ref{sec:cyclic}, the single bounce (dS or normal) of region {\bf 4} in section \ref{sec:single} and the asymmetric bounce of section \ref{sec:special}, where the bounce is not an extremum of temperature.

With the exception of string gas cosmology \cite{Brandenberger:2008nx}, early universe models have been heavily based on the scalar field realization, be it inflation or a bounce. Our consideration of Banks-Zaks theory therefore opens several new avenues in early universe research, which deserve further attention: \begin{enumerate}
    \item Using other field theories such as gauge theories for describing early universe evolution. We have taken a specific example, that is the Banks-Zaks theory near its conformal point. The crucial ingredient was the deviation from traceless energy momentum tensor with an anomalous dimension of the operator. It would be interesting if other CFTs or gauge theories slightly away from the conformal point can be as fruitful and provide other interesting results. 
    \item  At the macroscopic level, the thermal average of such theories is a huge simplification, as it allows to write down the energy density and pressure as functions of temperature only, even though they do not conform to the standard $p=w\rho$. Since we are interested in the global behavior of spacetime, this seems like a plausible simplification. One should consider the limitations of this approach. Specifically, an interesting analysis will be determining when will the microscopic degrees of freedom become relevant and the thermal average analysis looses its validity.
    \item The thermal average resulted in the temperature behaving as a time variable. This is not very surprising in Cosmology. However, to show the dynamics we did solve the equations using cosmic time. It will be nice to formulate the dynamics in terms of temperature only, as it may considerably simplify the analysis.
    \item The absence of a fundamental scalar field, immediately bypasses a multitude of theoretical questions such as Swampland issues \cite{Agrawal:2018own,Ben-Dayan:2018mhe,Artymowski:2019vfy}, and requires a redefinition and reanalysis of other questions like the stability of the theory, most notably in the case of NEC violation. When relevant, other questions such as eternal inflation and the measure problem have to be rephrased and assessed.
    \item Focusing on non-singular solutions, Banks-Zaks theory can easily be added to some other effective field theory describing inflation or a slow contraction, thus allowing to independently address the Big Bang singularity problem, regardless of the mechanism responsible for the observed CMB spectrum. Of specific interest is using it to provide the bounce needed in bouncing models. "Healthy" bouncing mechanisms, are scarce and in scalar field theories involve rather complicated lagrangians with non-canonical kinetic terms. This is due to the inherent instabilities usually associated with the NEC violation. Banks-Zaks theory is therefore a new bouncing mechanism, that could readily be combined with existing slow contraction or inflationary models.  Predictions of such combinations and possible distinct signatures should be calculated.
     \item As we have shown, the Banks-Zaks theory could support various early universe scenarios including inflation, slow contraction, exponential contraction and cyclic models. Hence, it allows for a multitude of valid background evolutions that solve the isotropy and flatness problem of the Hot Big Bang paradigm. For the sake of predictivity, an immediate question is its prediction for the scalar and tensor primordial spectrum. We are currently in the process of calculating these observables.
     \item For certain ranges of the parameter space, the unparticles seem to support possible Inflation or Dark Energy epochs, even without a bounce. We are currently analyzing these possibilities. 
\end{enumerate}
To summarize, Banks-Zaks theory in a cosmological background possesses a very reach phenomenology and our analysis has raised many novel issues for future analysis and discussion. 
\appendix
\section{Appendix - Different instabilities in the collapsing Universe}
In the appendix we wish to discuss the way various solutions reach a singularity at finite time or are unphysical for other reasons. These correspond to regions {\bf 2,3} and {\bf 5}.
As mentioned, whenever $p_0 >0$ one obtains a recollapse at $T = T_0$. The recollapse itself may be followed by a bounce (see Figs. \ref{fig:case1}, \ref{fig:case6}) or by a pole of $T$ and $H$ (see Fig. \ref{fig:case7}). We realize the source of the instability by noting that $H = \pm \sqrt{\rho/3}$ gives
\begin{equation}
\dot{x}=\pm \sqrt{\frac{\rho}{3}}\frac{a}{a_x} \, , \label{eq:Tdot}
\end{equation}
where $a_x = \frac{da}{dx}$. Let us consider the shrinking Universe (red and brown regions in the Fig. \ref{fig:alpha1}), which indicates the minus sign in \eqref{eq:Tdot}. For $\delta > 0${, $\delta>-3+1/w$} (region {\bf 2}) 
and $\alpha >\frac{\delta }{3 (\delta +4)}$ one finds in the big $x$ limit
\begin{equation}
x(t) \simeq \left(1-\frac{\sqrt{3 \alpha }}{2} (w+1) \left(\frac{3 (\alpha +1) (\delta +4)}{3 \alpha  (\delta +4)-\delta }\right)^{\frac{w+1}{2}} t \right)^{-\frac{2}{(\delta +3) (w+1)}} \, .
\end{equation}
Since, $1+w$ and $\delta$ are positive, $x(t)$ has a pole, which is a source of instability. 

For $\delta < 0$ we have to distinguish between red and brown regions. For $\dot{T}>0$ (red) one finds $x > 1$ after the recollapse. Thus, in the $x \gg 1$ limit one can obtain an analytical solution of \eqref{eq:Tdot}
\begin{eqnarray}
x(t) \simeq \sqrt{\frac{3}{3-2 \sqrt{3} t}} \quad \text{for} \quad w < \frac{1}{3} \, , \\ 
x(t) \simeq \left(1-\frac{ \sqrt{3\alpha}}{2}(w+1) \left(\frac{4 (\delta +3)}{\delta -3 \alpha  (\delta +4)}\right)^{\frac{w+1}{2}}t\right)^{-\frac{2}{3 (w+1)}} \quad \text{for} \quad w > \frac{1}{3} \, , \\
x(t) \simeq \left(1-\frac{2}{3} \sqrt{3 \left(\alpha  \left(\frac{4 (\delta +3)}{\delta -3 \alpha  (\delta +4)}\right)^{4/3}+1\right)} t \right)^{-\frac{1}{2}} \quad \text{for} \quad w = \frac{1}{3} \, .
\end{eqnarray}
For $-3\geq \delta \geq -6$, all of these solutions have a pole, which means that the temperature may obtain infinite values in finite time. Numerical precise calculations presented in left panel of Fig \ref{fig:case7} are in good agreement with the obtained approximate results. 

\begin{figure}[H]
\centering
\includegraphics[height=6cm]{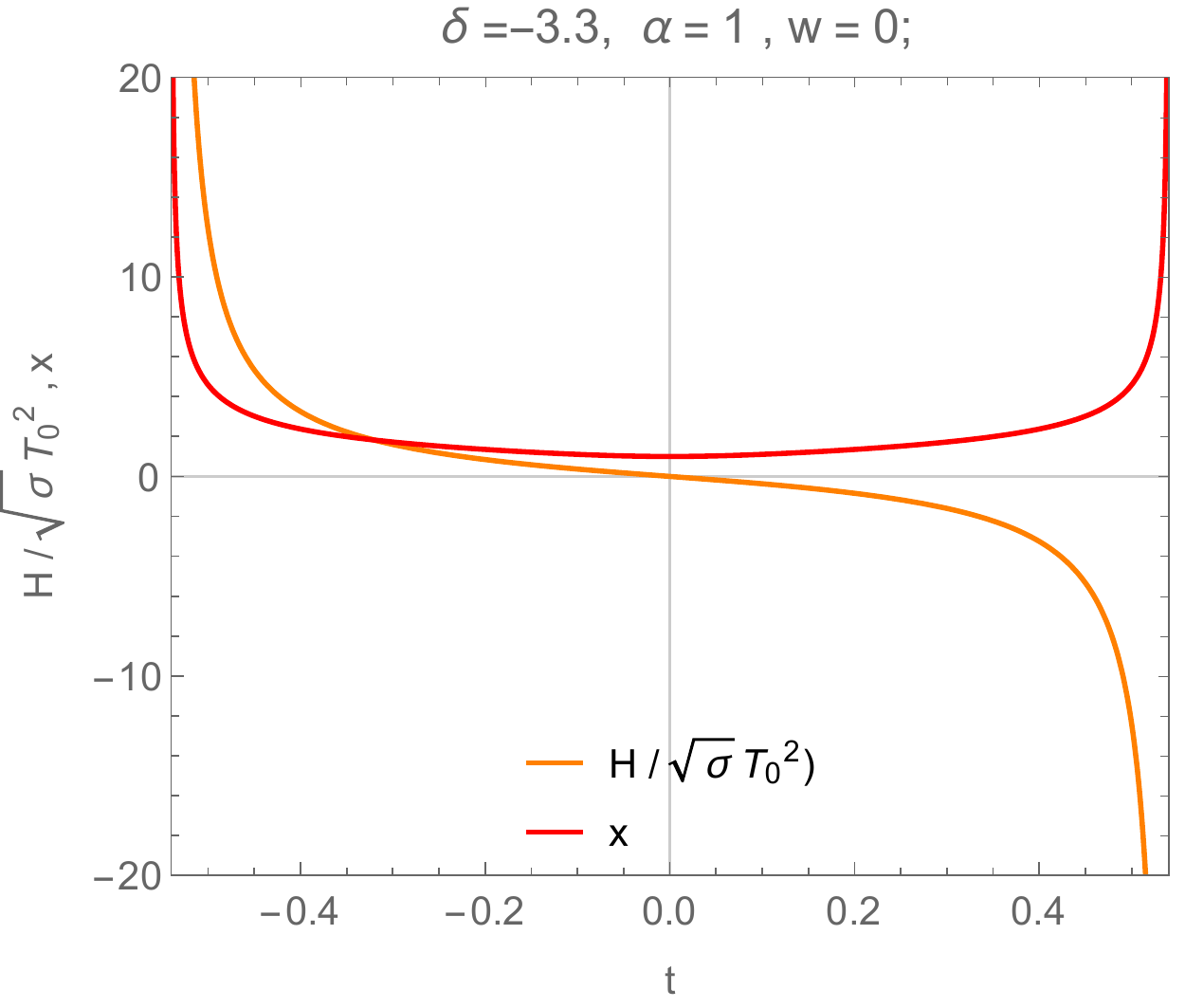} 
\includegraphics[height=6cm]{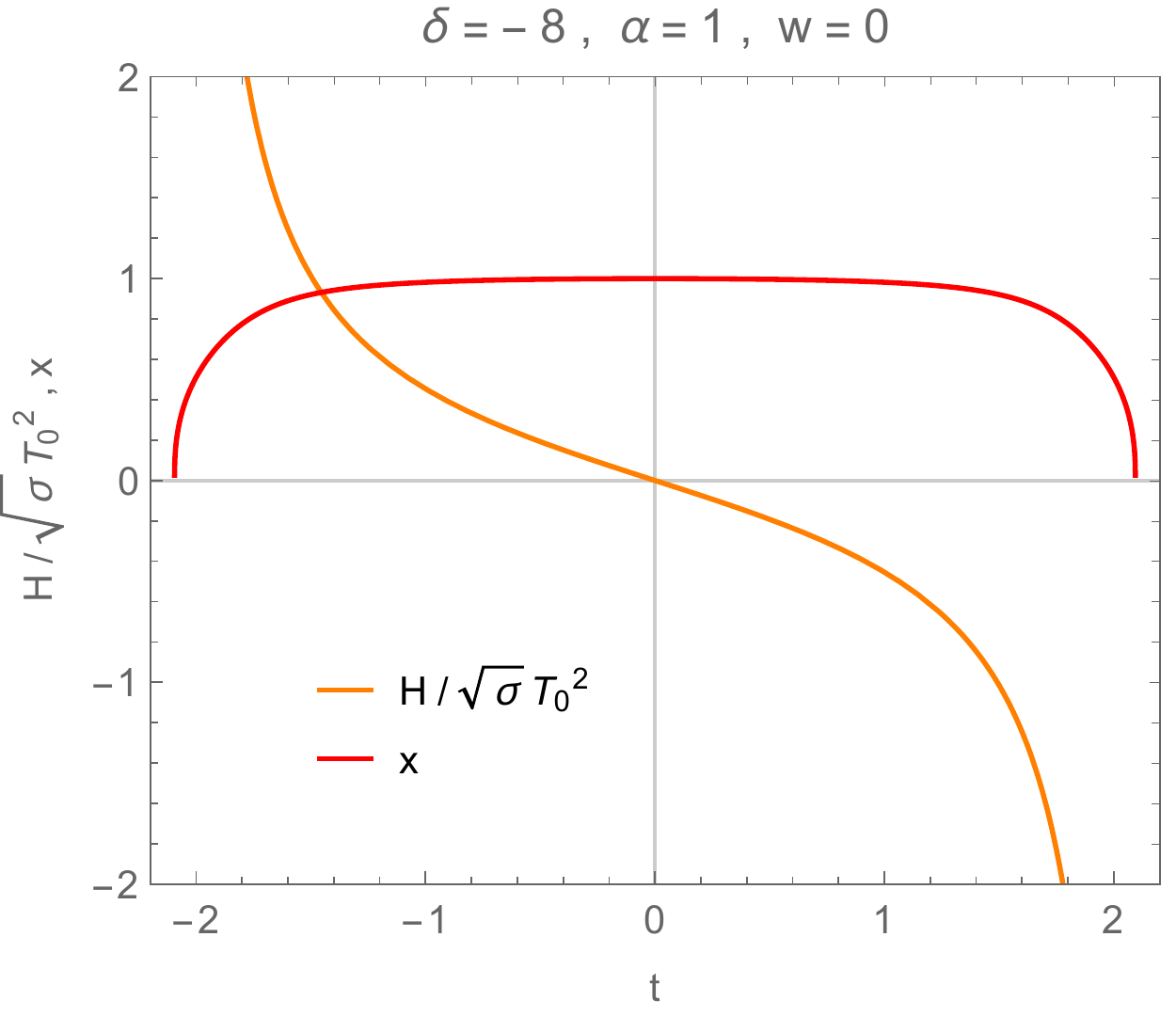}
\caption{\it Evolution of the normalized temperature $x=T/T_0$ (red) and the normalized Hubble parameter $ \frac{H}{\sqrt{\sigma} T_0^{2}}$ (orange). {Time is expressed in Planck units.} Left panel : The plot presents an example of the evolution in region {\bf 2} from Fig. \ref{fig:alpha1}. After the recollapse $x $ grows quickly until the solution becomes unstable. 
Right panel: The plot presents an example of region {\bf 5} from Fig. \ref{fig:alpha1}.  
The temperature obtains its maximum at the recollapse. Since $\delta < -4$, the $x ^{4+\delta}$ term from $\rho_u$ diverges, when $x \to 0$.} 
\label{fig:case7}
\end{figure}

For the recollapse with decreasing temperature (brown region) one obtains $\delta\leq -6$ and $x\leq 1$, which for small $x$ gives
\begin{equation}
x(t) \simeq \left(1-\frac{\sqrt{3 \alpha }}{2}(w+1) \left(\frac{3 (\alpha +1) (\delta +4)}{3 \alpha  (\delta +4)-\delta }\right)^{\frac{w+1}{2}} t \right)^{-\frac{2}{(\delta +3) (w+1)}} \, .
\end{equation}
Note that for certain $t$ one finds $x = 0$. Since $\rho_u$ contains a $x^{4+\delta}$ term and $4+\delta < 0$ for the brown region, one quickly obtains $\rho \to \infty$ for $x\to 0$. Right panel of Fig \ref{fig:case7} shows a numerical example of the evolution of the normalized Hubble parameter and temperature in region {\bf 5}.


\end{document}